 \documentclass[preprint2]{aastex}

\newtheorem{kam}{Theorem}[section]
\usepackage{graphicx}
 \usepackage{epsfig}
 \usepackage{amssymb}
\usepackage{amsmath}
 \usepackage{aas_macros}
\newcommand{\lgbya}{\log\frac{b}{a}}
\shorttitle{Normalization of hamiltonian \dots with Perturbations}
\shortauthors{Ram Kishor}

\begin{document}

%% LaTeX will automatically break titles if they run longer than
%% one line. However, you may use \\ to force a line break if
%% you desire.

\title{Normalization of Hamiltonian and Nonlinear Stability of the Triangular Equilibrium Points in Non-resonance Case with Perturbations}

% Use \author, \affil, and the \and command to format
%% author and affiliation information.
%% Note that \email has replaced the old \authoremail command
%% from AASTeX v4.0. You can use \email to mark an email address
%% anywhere in the paper, not just in the front matter.
%% As in the title, use \\ to force line breaks.

\author{Ram Kishor$^{1}$ and Badam Singh Kushvah$^{2}$ }
\affil{$^{1}$Department of Mathematics, Central University of Rajasthan, NH-8, Bandarsindari, Kishangarh, Ajmer-305817, Rajasthan, India}
\affil{$^{2}$Department of   Applied  Mathematics,
Indian School of Mines, Dhanbad - 826004, Jharkhand, India}
% 
% \author{C. D. Biemesderfer\altaffilmark{4,5}}
% \affil{National Optical Astronomy Observatories, Tucson, AZ 85719}
\email{$^{1}$kishor.ram888@gmail.com; $^{2}$bskush@gmail.com} 

\begin{abstract}
For the study of nonlinear stability of a dynamical system, normalized Hamiltonian of the system is very important to discuss the dynamics in the vicinity of  invariant objects. In general, it represents a nonlinear approximation to the dynamics, which is very helpful to obtain the information about realistic solution of the problem. Present paper reflects about normalization of the Hamiltonian and analysis of nonlinear stability in non-resonance case, in the  Chermnykh-like problem under the influence of perturbations in the form of radiation pressure, oblateness, and a disc. To describe nonlinear stability, initially, quadratic part of the Hamiltonian is normalized in the neighborhood of triangular equilibrium point and then higher order normalization is performed. Due to the presence of perturbations and a tedious huge algebraic computation for intermediate terms, we have computed only up to the fourth order normalized Hamiltonian using Lie transforms. In non-resonance case, nonlinear stability of the system is discussed with the help of Arnold-Moser theorem. Again, the effects of radiation pressure, oblateness and presence of the disc are analyzed, separately and it is observed that in the absence as well as presence of perturbation parameters, triangular equilibrium point is unstable in nonlinear sense within the stability range $0<\mu<\mu_1=\bar{\mu_c}$ due to failure of Arnold-Moser theorem. However, perturbation parameters affect the values of $\mu$ at which $D_4=0$, significantly. This study may help to analyze more generalized cases of the  problem in the presence of some other types of perturbations such as P-R drag and solar wind drag.  The results are limited to the regular symmetric disc but in future it can be extended.
\end{abstract}


\section{Introduction}
\label{sec:nh}
Study of stability property of a dynamical system is a necessary step which brings not only the system to tackle many realistic problems of the world but also helps to understand the motion of test particle for a long time of evolution. The stability of the system for a long time of evolution is an important and critical issue and hence, a number of researchers are studying the Hamiltonian system of the problem in the vicinity of elliptic equilibrium point in many fields such as mathematical physics, dynamical astronomy, astronomy, celestial mechanics etc.  

Many researchers \citep{Deprit1967AJ.....72..173D, Markeev1977SvA....21..507M, Mayer, coppola1989ZaMM...69..275C, K.Gozdzieski1998CeMDA..70...41G,Jorba} have studied restricted problem of three bodies in the context of stability in classical cases and some of the researchers  \citep{Bhatnagar1983CeMec..30...97B,Markellos1996Ap&SS.245..157M,Ishwar1997CeMDA..65..253I,SubbaRao1997CeMDA..65..291S,Kushvah2007Ap&SS.312..279K,Alvarez-Ram2012arXiv1212.2179A} have discussed the  stability for generalized cases.  However, a very little attention has been given to the  problem  with the effect of perturbations such as radiation pressure, oblateness, drag forces, and presence of a disc like structure in the problem. In the present paper, we consider  Chermnykh-like problem under the influence of perturbations in the form of radiation pressure, oblateness and presence of a disc, which is rotating about common center of mass of the system. Chermnykh-like problem is a result of some modification in original Chermnykh's problem which consists with the motion of a point mass in a plane under the influence of gravitational effect of a uniformly rotating dumb-bell and it was first time studied by \citet{Chermnykh1987VeLen.......73C}. This problem has a number of applications in different areas such as celestial mechanics, dynamical astronomy, extra solar planetary system and chemistry \citep{Gozdziewski1999CeMDA..75..251G,Rivera2000ApJ...530..454R,Jiang2001AA...367..943J,Jiang2004IJBC...14.3153J,Jiang2004astro.ph..4408J}.  The different aspects of the problem such as existence of equilibrium points,  stability analysis in resonance and non-resonance cases, computation of orbits, Lyapunov characteristic exponent of trajectories etc. have been studied by many authors \citep{K.Gozdzieski1998CeMDA..70...41G,Gozdziewski1999CeMDA..75..251G,Gozdziewski2003CeMDA..85...79G,Papadakis2005Ap&SS.299..129P,Papadakis2005Ap&SS.299...67P,Jiang2006Ap&SS.305..341J,YehLCJiang2006,Kushvah2008Ap&SS.318...41K,Kushvah2012Ap&SS.337..115K,Kishor2013P&SS...84...93K}.  \cite{Gabern2005Nonli..18.1705G} studied KAM stability of Trojan asteroid under the frame of planar restricted three body problem and \cite{benettin1998nekhoroshev} described condition of the applicability of the Nekhoroshev stability theorem. Moreover, nonlinear stability of Trojan asteroid in the sense of Nekhoroshev stability has described by many researcher \citep{LittlewoodPLMS:PLMS0343,LittlewoodPLMS:PLMS0525,Giorgilli1997A&A...317..254G,Efthymiopoulos2005MNRAS.364..253E,Lhotka2008MNRAS.384.1165L}. In the present paper, we are interested to discuss nonlinear stability of triangular equilibrium point in non-resonance case under the influence of perturbations in the form of radiation pressure, oblateness and the disc with the help of Arnold-Moser theorem \citep{Mayer, K.Gozdzieski1998CeMDA..70...41G}. 

In order to discuss, nonlinear stability of triangular equilibrium point in non-resonance case with the help of Arnold- Moser theorem, we  obtain normal forms of the Hamiltonian of the system up to a finite order, which are very important to discuss the dynamics in the neighborhood of invariant objects. Several researchers \citep{poincarea,poincarec,brikhoff,Deprit1969CeMec...1...12D,takensa,takensb,ushikia,ushikib,Jorba,coppola1989ZaMM...69..275C}  have described the different  normalization processes and also, they have utilized the normalized Hamiltonian to analyze the nonlinear stability of the dynamical system. The main idea behind the normal form is to construct a suitable transformation of phase space which yields the simplest form up to a certain order of accuracy of a given system of differential equations. In short, it can be used to approximate the dynamics and hence, study of real world problems. 
There are several approaches  \citep{brikhoff, Deprit,takensa, takensb, dragtfinn, ushikib} to find the transformation equations to reduce the Hamiltonian into  simplest form. We have performed the normalization of Hamiltonian of the system up to fourth order  by the method of Lie transforms which are described well in \cite{Jorba} and \cite{coppola1989ZaMM...69..275C}.

The paper is organized as follows: Section-\ref{sec:form} described the formulation of problem whereas, diagonalization of the Hamiltonian is discussed in detail under Section-\ref{sec:norm1}. Section-\ref{sec:nstb} is devoted to non-linear stability in non-resonance case by the use of Arnold-Moser theorem, whereas Subsection-\ref{subsec:norm2} contains computation of coefficients of the normalized Hamiltonian  up to order four on the basis of Lie transform.  Finally, the results are concluded in Section-\ref{sec:con}. Algebraic as well as numerical computation has been  performed with the help of Mathematica$^{\textregistered}$ \citet{wolfram2003mathematica} software package.

\section{Formulation of the problem}
\label{sec:form}
Mathematical formulation of the problem is similar to \cite{Kushvah2012Ap&SS.337..115K} whereas, for self sufficient paper it is as follows. We consider the motion of infinitesimal mass under the influence of gravitational field of massive bodies (also known as primaries, here bigger primary is taken as radiating body and smaller is an oblate spheroid) and perturbations in the form of radiation pressure of bigger primary, oblateness  of smaller primary and a disc, which is rotating about the common center of mass of the system having power-law density profile $\rho(r)=\dfrac{c}{r^{p}}$, where $p\in \mathbb{N}$ (here, we have taken $p=3$) and $c$ is a constant which depends on total mass of the disc. It is assumed that the effect of infinitesimal mass on the motions of both the primaries as well as of the disc, is negligible. The proposed model can be realize by considering, a disc about the common center of mass of Sun-Planet system and an infinitesimal body such as spacecraft or satellite moves under the influence of celestial forces. Units of mass and distance are taken as the  sum of masses of the primaries and separation between them, respectively whereas, unit of time is taken as time period of rotating frame. 

Under these assumption, Hamiltonian function  of the  Chermnykh-like problem in the presence of radiation pressure, oblateness and the disc, in the phase coordinate $(x, \, y, \, p_x, \, p_y)$, is written as \citep{Kishor2013P&SS...84...93K}:
\begin{eqnarray}
H\left(x,y,p_x,p_y\right)&=&\frac{1}{2}\left(p^2_x+p^2_y\right)+\mathbf{n}\left(yp_x-xp_y\right)\nonumber\\&&-\frac{(1-\mu)q_1}{r_1}-(1+\frac{A_2}{2r_{2}^{3}})\frac{\mu}{r_2}\nonumber\\&&-\pi c h\left[\frac{2(b-a)}{abr}+\frac{7\lgbya}{8r^2}\right],\label{eq:hf}
\end{eqnarray}
where  $p_x=\dot{x}-\mathbf{n}y$ and  $p_y=\dot{y}+\mathbf{n}x$ are momenta coordinate. Last term on the right side is due to presence of the disc. Mean motion of the system is given as:
\begin{eqnarray}
\mathbf{n}&=&\sqrt{q_1+\frac{3}{2}A_2-2f_b(r)},\label{eq:mm}\end{eqnarray}
 where mass reduction factor
$q_1=(1-\frac{F_p}{F_g})$ \citep{Ragos1993Ap&SS.209..267R} with $F_p$  and $F_g$ as the radiation pressure and gravitational attraction forces respectively (here, $0< q_1< 1$ because in solar planetary system for radiating body as the Sun, $\frac{F_p}{F_g}< 1$)\citep{schuerman1980ApJ...238..337S}; oblateness coefficients $A_2=\frac{R^{2}_e-R^{2}_p}{5R^{2}}$ \citep{McCuskey1963QB351.M3}, with $R_e$ and $R_p$ be the equatorial and polar radii of the oblate body, respectively and $R$ is the separation  between both the primaries (here, $0< A_2< 1$ for oblate body but for prolate body $-1< A_2< 0$); $f_b(r)$ is the gravitational force due to the disc which is given as \citep{Kushvah2012Ap&SS.337..115K}
\begin{eqnarray}
f_b(r)&=&-\pi c h\left[\frac{2(b-a)}{abr}+\frac{7\lgbya}{8r^2}\right] ,\label{eq:fbr} 
\end{eqnarray} where $a, b$ are inner and outer radii respectively, of the radially symmetric disc (here, the dimension of the disc is taken in such a way that disc width $b-a< 0.3$ whereas, thickness of the disc is $h=10^{-4}$ and constant $c=1910.83$).  $\mu=\frac{m_J}{M_S+m_J}$ be the mass parameter in the Sun-Jupiter system ($M_S$ and $m_J$ are masses of the Sun and the Jupiter, respectively). 

The coordinates of triangular equilibrium points of the problem are given as \citep{Kushvah2012Ap&SS.337..115K}:
\begin{eqnarray} 
x_e&=&\frac{q_1^{\frac{2}{3}}}{2}-\mu+(q_1^{\frac{2}{3}} \delta_1-\delta_2),\label{eq:xx}\\ 
y_e&=&\pm q_1^{\frac{1}{3}}[1-\frac{q_1^{\frac{2}{3}}}{4}+(2-q_1^{\frac{2}{3}}) \delta_1+\delta_2]^{\frac{1}{2}}, \label{eq:yy}
 \end{eqnarray}
 where 
 \begin{eqnarray}
\delta_1&=&\frac{1}{3}\left[1-\mathbf{n}^2 +\frac{2 \pi c h (b-a)}{a b \{\mu^2 +q_1^{\frac{2}{3}}(1-\mu)\}^{\frac{3}{2}}}\right.\nonumber\\&&\left.+\frac{3 \pi c h {\lgbya}}{8\{\mu^2 +q_1^{\frac{2}{3}}(1-\mu)\}^2}\right], \label{eq:dl1} \\
\delta_2&=&\frac{1}{3(1+\frac{5}{2}A_2)}\left[1-\mathbf{n}^2+\frac{3A_2}{2}+\right.\nonumber\\&&\left.\frac{2\pi c h(b-a)}{a b\{\mu^2 +q_1^{\frac{2}{3}}(1-\mu)\}^{\frac{3}{2}}}+\frac{3 \pi c h{\lgbya}}{8\{\mu^2 +q_1^{\frac{2}{3}}(1-\mu)\}^2}\right]\label{eq:dl2}\end{eqnarray} 
are small quantities.
\section{Diagonalization of Hamiltonian}
\label{sec:norm1}

For the simplicity and to over come the expression computation's complexity, we have considered only linear order terms in perturbing parameters through the computations in the paper. Therefore, before diagonalization of the Hamiltonian, we obtain mean motion and hence, triangular equilibrium points  $L_{4,5}(x_e, \, y_e)$
 as a linear function of parameters $\mu$, \, $q_1, \, A_2$ and $b$. Since, $q_1<1, \, A_2<1$ 
and $b>1$ so, we have supposed that  $q_1=1-\epsilon_{1}$ and $b=1+\epsilon_{2}$, 
$0<\epsilon_{1}, \epsilon_{2}\ll1$. First, we have expanded mean motion $(\mathbf{n})$ and then coordinates of triangular equilibrium points $(x_e, \, y_e)$ 
about $\mu=0, \, \epsilon_{1}=0, \, A_2=0$ and $\epsilon_{2}=0$, respectively and finally, 
taking linear order terms of $\mu, \, \epsilon_{1}, \, A_2$ and $\epsilon_{2}$, we 
get 
\begin{eqnarray}
 \mathbf{n}&=& 1+\frac{3 A_2}{4}-\frac{\epsilon_1}{2}+\frac{3 \epsilon_2}{2}\label{eq:n},\\
x_e&=&\frac{1}{2}-\mu-\frac{\epsilon_1}{3},\\
y_e&=&\pm \frac{\sqrt{3}}{2}\left( 1-\frac{2A_2}{3}+\frac{\epsilon_1}{4}-\frac{2\epsilon_2}{3}\right),
\end{eqnarray}
where $\textquoteleft +\textquoteright$ sign corresponds to $L_4$ point and 
$\textquoteleft -\textquoteright$ for $L_5$. We have discussed the stability of $L_4$ 
point whereas, dynamics of $L_5$ is similar to that of $L_4$. For the convenience, we shift 
the origin at equilibrium point $L_4$ using simple translation.
\begin{eqnarray}
 &&x^{*}=x-x_e, \quad y^{*}=y-y_e, \nonumber\\&& p^{*}_{x}=p_{x}+y_e, \quad p^{*}_{y}=p_{y}-x_e.
\end{eqnarray}
Applying this change of variable to the Hamiltonian (\ref{eq:hf}), we obtain
\begin{eqnarray}
H^{*}&=&\frac{1}{2}\left[(p^{*}_{x}-y_e)^2+(p^{*}_{y}+x_e)^2\right]+\nonumber\\&& n\left[(y^{*}+y_e)(p^{*}_{x}-y_e)-(x^{*}+x_e)(p^{*}_{y}+x_e)\right]\nonumber\\&&
-\frac{(1-\mu)q_1}{[(x^{*}+x_e+\mu)^2+(y^{*}+y_e)^2]^{\frac{1}{2}}}\nonumber\\&&-\frac{\mu}{(x^{*}+x_e+\mu-1)^2+(y^{*}+y_e)^2]^{\frac{1}{2}}}\nonumber\\&&
-\frac{\mu A_2}{2[(x^{*}+x_e+\mu)^2+(y^{*}+y_e)^2]^{\frac{3}{2}}}\nonumber\\&&-\pi c h\left[\frac{2(b-a)}{ab[(x^{*}+x_e)^2+(y^{*}+y_e)^2]^{\frac{1}{2}}}\right.\nonumber\\&&\left.+\frac{7\lgbya}{8[(x^{*}+x_e)^2+(y^{*}+y_e)^2]}\right].\label{eq:hfdc}
\end{eqnarray}
Expanding the resulting  Hamiltonian in Taylor 
series about origin (which is actually the triangular equilibrium point), as follows
\begin{eqnarray}
&&H^*=H_0+H_1+H_2+H_3+H_4+\dots,\label{eq:tehm1}
\end{eqnarray} where 
\begin{eqnarray}
&&H_n=\sum{H_{jkls}}{x^{*}}^{j}{y^{*}}^{k}{p^{*}_{x}}^{l}{p^{*}_{y}}^{s}\quad \text {with}\nonumber\\&&j+k+l+s=n \label{eq:tehm2}.
\end{eqnarray}Since, origin is an equilibrium point therefore, first order term $H_1$ must 
vanish whereas, constant term $H_0$ drop out because it is irrelevant to the 
dynamics. The quadratic term $H_2$, is useful for higher order normal forms, around the triangular 
point $L_4$, and given as
\begin{eqnarray}
H_2&=&\frac{1}{2}\left({p^{*}_{x}}^{2}+{p^{*}_{y}}^{2}\right)+\mathbf{n}\left(y^*p^{*}_{x}-x^*p^{*}_{y}\right)-\nonumber\\&&\frac{1}{2}\left(P {x^{*}}^2 +Q {y^{*}}^2+S x^{*}y^{*}\right),\label{eq:h2f}
\end{eqnarray}  where coefficients $P, \, Q$ and $S$ are given as
\begin{eqnarray}
&&P=-\frac{1}{4}\left( 1+\frac{49\epsilon_1}{16}-\frac{3A_2}{2}-\frac{\epsilon_2}{4}\right),\label{eq:Pc}\\&&
Q=\frac{5}{4}\left( 1-\frac{19\epsilon_1}{80}+\frac{9A_2}{10}+\frac{5\epsilon_2}{4}\right),\label{eq:Qc}\\&&
S=\frac{3\sqrt{3}(1-2\mu)}{4}\left( 1+\frac{73\epsilon_1}{48}+\frac{11A_2}{6}-\frac{53\epsilon_2}{12}\right).\label{eq:Sc}
\end{eqnarray}
Thus, $H_2$ becomes
\begin{eqnarray}
H_2&=&\frac{1}{2}\left({p^{*}_{x}}^{2}+{p^{*}_{y}}^{2}\right)+\mathbf{n}\left(y^*p^{*}_{x}-x^*p^{*}_{y}\right)\nonumber\\&&-\frac{1}{2}\left(\frac{-1}{4} {x^{*}}^2 +\frac{5}{4} {y^{*}}^2+a_0 x^{*}y^{*}\right)\nonumber\\&&-\frac{1}{2}\left(\frac{-49}{64} {x^{*}}^2 -\frac{19}{64}{y^{*}}^2+a_{1} x^{*}y^{*} \right)\epsilon_{1}\nonumber\\&&-\frac{1}{2}\left(\frac{3}{8} {x^{*}}^2 +\frac{9}{8} {y^{*}}^2+a_{2} x^{*}y^{*}\right)A_2\nonumber\\&&-\frac{1}{2}\left(\frac{1}{16} {x^{*}}^2 +\frac{25}{16} {y^{*}}^2+a_{3} x^{*}y^{*}\right)\epsilon_2,\label{eq:exh2f}
\end{eqnarray}where
\begin{eqnarray}
&&a_0=\frac{3\sqrt{3}(1-2\mu)}{4}, \, a_1=\frac{73\sqrt{3}(1-2\mu)}{64}, \nonumber\\&&
a_2=\frac{11\sqrt{3}(1-2\mu)}{8}, \, a_3=\frac{-53\sqrt{3}(1-2\mu)}{16}\end{eqnarray}
Since,  we are dealing the problem in the presence of three type of perturbations in the form of radiation pressure, obletness and the disc, therefore, for simplicity,  coefficients $H_{jkls}$ in  equation (\ref{eq:tehm2}) are splitted into four parts such as $g_{jkls}, \, g_{jklse1}$,  $g_{jklsA}$ and $g_{jklse2}$, which correspond to the terms due to classical case (i.e. absence of perturbations),  radiation effects $\epsilon_1$, oblateness $A_2$ and presence of the disc $\epsilon_2$, respectively. In other words,
\begin{eqnarray}
H_{jkls}=g_{jkls}+g_{jklse1}+g_{jklsA}+g_{jklse2}\label{eq:H0jkls}, 
\end{eqnarray}where $g_{jkls}$ represents coefficients for classical part, $g_{jklse1}$ indicates coefficients  for radiation pressure terms, $g_{jklsA}$ used for oblateness and $g_{jklse2}$ corresponds to the disc with $j, \, k, \, l, \, s=0, \, 1, \, 2,\,3,\,4$ such that $j+k+l+s=4$. However, in the absence of perturbations i.e. when $A_2=\epsilon_1=\epsilon_2=0$ then  $H_{jkls}=g_{jkls}$,  which is equivalent to the coefficients of the Hamiltonian in classical case.

Since, Hamilton's equations of motion of the infinitesimal mass are written as
\begin{eqnarray}\begin{bmatrix}
\dot{x^{*}}\\
\dot{y^{*}}\\
\dot{p^{*}_x} \\
\dot{p^{*}_y} 
\end{bmatrix}=J_4.\nabla H_2= J_4. Hess[H_2]\begin{bmatrix}
x^*\\
y^*\\
p^{*}_x \\
p^{*}_y 
\end{bmatrix},
\label{eq:xt}\end{eqnarray}
where 
\begin{eqnarray}
J_4=\begin{bmatrix}
0&0&1&0 \\
0&0&0&1 \\
-1&0&0&0\\
0&-1&0&0
\end{bmatrix},
\label{eq:J}\end{eqnarray} and
\begin{eqnarray}J_4.Hess[H_2]=\mathbf{M}=\begin{bmatrix}
0&\mathbf{n}&1&0 \\
-\mathbf{n}&0&0&1 \\
P&S&0&\mathbf{n}\\
S&Q&-\mathbf{n}&0
\end{bmatrix}.
\label{eq:hessM}\end{eqnarray}
 The characteristic equation of the monodromy matrix $\mathbf{M}$ is 
\begin{eqnarray}
&& \lambda^4+\left(-P^2-Q^2+2\mathbf{n}^2\right)\lambda^2+\left(\mathbf{n}^2P+\mathbf{n}^2Q\right.\nonumber\\&&\left.+PQ-S^2+\mathbf{n}^4\right)=0.\label{eq:Hce}
\end{eqnarray}
From equation (\ref{eq:Hce}), it can be easily obtained that system (\ref{eq:xt}) 
is stable, if the mass parameter $\mu$ satisfy the condition $0<\mu<\bar\mu_c$, where  $\bar\mu_c$ is the Routh value of 
the mass ratio of the problem \citep{Kishor2013MNRAS.436.1741K}. Since, we are studying the same case $0<\mu<\bar\mu_c$ so, it is 
assumed that four roots of characteristic equation (\ref{eq:Hce}) are purely 
imaginary say, $\lambda_{1,2}=\pm i \omega_1$ and $\lambda_{3,4}=\pm i \omega_2$. 
As, the real values of $\omega_{1,2}$ are frequencies of the linear oscillations 
of the infinitesimal mass at the equilibrium point $L_4$ and it is obvious that 
they differ for the stability region $0<\mu<\bar\mu_c$.  

Now, our aim is to obtain a real symplectic change of variable due to which one 
can find the real diagonalize  Hamiltonian from (\ref{eq:h2f}). For 
that, first step is to obtain the characteristic vectors of the matrix $\mathbf{M}$ 
corresponding to the characteristic roots. If we denote the matrix $\mathbf{M-\lambda 
I_4}$ by $\mathbf{M_\lambda}$ \citep{Jorba}, then
\begin{eqnarray}\mathbf{M_\lambda}=\begin{bmatrix}
\mathbf{M_\lambda^{'}}&\mathbf{I_2} \\
\mathbf{M^{'}}&\mathbf{M_\lambda^{'}}
\end{bmatrix},\label{eq:egv1}\end{eqnarray}
where \begin{eqnarray}
\mathbf{M_\lambda^{'}}=\begin{bmatrix}
-\lambda&\mathbf{n} \\
-\mathbf{n}&-\lambda
\end{bmatrix}, \quad \mathbf{M^{'}}=\begin{bmatrix}
P&S \\
S&Q
\end{bmatrix}  
\label{eq:egv2}\end{eqnarray}
and $\mathbf{I_2}$ is identity matrix of $2\times2$.
Since, $\lambda$ is a root of the matrix $\mathbf{M}$ and hence, the kernel of 
$\mathbf{M_\lambda}$ is obtained by solving the system
\begin{eqnarray}\begin{bmatrix}
\mathbf{M_\lambda^{'}}&\mathbf{I_2} \\
\mathbf{M^{'}}&\mathbf{M_\lambda^{'}}
\end{bmatrix}.\begin{bmatrix}
\mathbf{X_1} \\
\mathbf{X_2}
\end{bmatrix}=\begin{bmatrix}
0 \\
0
\end{bmatrix},  
\label{eq:ker}\end{eqnarray}where $\,\mathbf{X_1}=\begin{bmatrix}
x^* \\
y^*
\end{bmatrix}$ and $\mathbf{X_2}=\begin{bmatrix}
p^{*}_x \\
p^{*}_y
\end{bmatrix}$
i.e.
\begin{eqnarray}
&&\mathbf{M_\lambda^{'} X_1}+\mathbf{I_2 X_2}={0},\label{eq:ker1}\\&&
  \mathbf{M^{'} X_1}+\mathbf{M_\lambda^{'} X_2}={0}.\label{eq:ker2}\end{eqnarray}
From above equations, we have
\begin{eqnarray}\begin{bmatrix}
\mathbf{n}^2+P-\lambda&S+2\mathbf{n}\lambda \\
S+2\mathbf{n}\lambda&\mathbf{n}^2+Q-\lambda^2
\end{bmatrix}.\begin{bmatrix}
x^* \\
y^*
\end{bmatrix}=\begin{bmatrix}
0 \\
0
\end{bmatrix},  
\label{eq:ker3}\end{eqnarray}
which gives either 
\begin{eqnarray}&&x^*=S+2\mathbf{n}\lambda, \, y^*=-(\mathbf{n}^2+P-\lambda^2)\label{eq:chvtr1} 
\quad\\
\text{or} &&x^*=(\mathbf{n}^2+Q-\lambda^2), \, 
y^*=-S+2\mathbf{n}\lambda.\label{eq:chvtr2}\end{eqnarray}
 If we take first set (\ref{eq:chvtr1}) of $\,x^*\,$ and $\,y^*\,$, then final symplectic 
matrix is similar to that of  \cite{Jorba}, whereas if we use second set (\ref{eq:chvtr2}) 
of  $\,x^*\,$ and $\,y^*\,$, results agree with that of \cite{Deprit1967AJ.....72..173D} and \cite{Mayer}.

Here, we use second set (\ref{eq:chvtr2}) of$x^*$ and $y^*$ to proceed further. Putting, these $x^*$ and $y^*$ into equation (\ref{eq:ker2}), we get
\begin{eqnarray}&&p^{*}_x=-\lambda^3-(\mathbf{n}^2-Q)\lambda+\mathbf{n}S,\label{eq:chvtr3}\\&&
 p^{*}_y=\mathbf{n}\lambda^2-S\lambda+\mathbf{n}Q+\mathbf{n}^3.\label{eq:chvtr4}\end{eqnarray}
Thus, from the equations (\ref{eq:chvtr2}-\ref{eq:chvtr4}), the characteristic vector of the matrix $\mathbf{M}$ is given as 
\begin{eqnarray}\begin{bmatrix}
x^*\\
y^*\\
p^{*}_x \\
p^{*}_y
\end{bmatrix}=\begin{bmatrix}
(\mathbf{n}^2+Q-\lambda^2)\\
-S+2\mathbf{n}\lambda\\
-\lambda^3-(\mathbf{n}^2-Q)\lambda+\mathbf{n}S\\
\mathbf{n}\lambda^2-S\lambda+\mathbf{n}Q+\mathbf{n}^3 
\end{bmatrix}.
\label{eq:cv}\end{eqnarray}

Since, the characteristic roots of the matrix are pure imaginary and given as 
$\lambda=i\omega, \, \omega\in\mathbb{R}$, which can be obtained  with 
the help of equation
\begin{eqnarray}
 &&\omega^4-\left(-P^2-Q^2+2\mathbf{n}^2\right)\omega^2+\left(\mathbf{n}^2P+\mathbf{n}^2Q\right.\nonumber\\&&\left.+PQ-S^2+\mathbf{n}^4\right)=0,\label{eq:wce}
\end{eqnarray}
which provide $\lambda_{1,2}=\pm i\omega_{1}$ and  $\lambda_{3,4}=\pm i\omega_{2}$. The frequencies $\omega_{1,2}$ in terms of $\epsilon_1, \, A_2, \, \epsilon_2$ and $\mu$, are obtained with the help of equations (\ref{eq:n}), (\ref{eq:Pc}-\ref{eq:Sc}) and (\ref{eq:wce}) under linear approximation of $\epsilon_1, \, A_2, \, \epsilon_2$, and these are given as
\begin{eqnarray}
 &&\omega_1=\left[\frac{B_1+\sqrt{B_{1}^2-4B_2}}{2}\right]^{\frac{1}{2}},\label{eq:w1}\\
&&\omega_2=\left[\frac{B_1-\sqrt{B_{1}^2-4B_2}}{2}\right]^{\frac{1}{2}}\label{eq:w2}
\end{eqnarray} with
\begin{eqnarray}
 &&B_1=1-\frac{15\epsilon_1}{16}+\frac{3A_2}{2}+\frac{35\epsilon_2}{8},\label{eq:B1}\\
&&B_2=\frac{27}{16}-\frac{27(1-2\mu)^2}{16}-\frac{633\epsilon_1}{128}\nonumber\\&&+\frac{99A_2}{16}+\frac{165\epsilon_2}{16}.\label{eq:B2}
\end{eqnarray} 
Again, if we put $\lambda=i\omega$ into characteristic vector (\ref{eq:cv}) and 
then separating real and imaginary parts, say $u$ and $v$, of resulting 
characteristic vectors, then we obtain

\begin{eqnarray}u=\begin{bmatrix}
\mathbf{n}^2+Q+\omega^2\\
-S\\
\mathbf{n}S\\
-\mathbf{n}\omega^2+\mathbf{n}Q+\mathbf{n}^3 
\end{bmatrix},
\label{eq:wr}\end{eqnarray}
\begin{eqnarray}v=\begin{bmatrix}
0\\
2\mathbf{n}\omega\\
-\mathbf{n}^2\omega+Q\omega+\omega^3\\
-S\omega 
\end{bmatrix}.
\label{eq:wi}\end{eqnarray}
Now, consider the required symplectic change of phase variable  is given 
by the matrix $\mathbf{C}=(v_1, \, v_2, \, u_1, \, u_2)$, where $u_j, \, v_j, \, 
j=1,2$ represent the values of $u, \, v$ correspond to frequencies $\omega_j, \, 
j=1,2$, respectively. Thus, it is obvious the symplectic change satisfy the property $\mathbf{C}^{T}\mathbf{J_4}\mathbf{C}=\mathbf{J_4}$. Substituting  
\begin{eqnarray}
 &&\omega^2_{1}+\omega^2_{2}=2\mathbf{n}^2-P-Q,\label{eq:wsum}\\
&&\omega^2_{1}\omega^2_{2}=\mathbf{n}^4+\mathbf{n}^2(P+Q)+PQ-S^2\label{eq:wpdt},
\end{eqnarray} and simplifying, we obtain
\begin{eqnarray}\mathbf{C}^{T}\mathbf{J_4}\mathbf{C}=\begin{bmatrix}
\mathbf{0}&\mathbf{D} \\
-\mathbf{D}&\mathbf{0}
\end{bmatrix},\quad \mathbf{D}=\begin{bmatrix}
d(\omega_1)&0 \\
0&d(\omega_2)
\end{bmatrix},  
\label{eq:ctjc}\end{eqnarray}
where 
\begin{eqnarray}
 d(\omega)&=&2\omega\left[2\omega^4+(P+3Q)\omega^2+PQ+Q^2\right.\nonumber\\&&\left.+\mathbf{n}^2(P-Q-2\mathbf{n}^4)\right].\label{eq:dw}
\end{eqnarray}
Since, in order to satisfy the symplectic property, $d(\omega)$ should be one, 
if it is not then we need to scale the columns of $\mathbf{C}$ by the 
quantity $\sqrt{d(\omega_j)}, \, j=1,2$ i.e. 
\begin{eqnarray*}&&\mathbf{C}=\left(\frac{v_1}{\sqrt{d(\omega_1)}}, \, \frac{v_2}{\sqrt{d(\omega_2)}}, \, \frac{u_1}{\sqrt{d(\omega_1)}}, \, \frac{u_2}{\sqrt{d(\omega_2)}}\right).\end{eqnarray*}
 Now, this matrix is symplectic but, we require real symplectic change, so it is necessary 
to take $d(\omega_j)>0, \, j=1,2$ which is possible when we take $\omega_1>0$ 
and $\omega_2<0$. Thus, the transformation obtained is real and 
symplectic and gives diagonalize form of the Hamiltonian (\ref{eq:h2f}) as
\begin{eqnarray}
 &&H_2=\frac{\omega_{1}}{2}({\mathbf{x}}^2+{\mathbf{p_{x}}}^2)+\frac{\omega_{2}}{2}({\mathbf{y}}^2+{\mathbf{p_{y}}}^2).\label{eq:nh2f}
\end{eqnarray}
In order to solve homological equations \citep{Jorba}, which determine 
the generating function, in an easier way, we have to change the normalized 
Hamiltonian (\ref{eq:nh2f}) into complex normal form with the help of an other 
symplectic change of variable, which are given as follows:
\begin{eqnarray}
 &&\mathbf{x}=\frac{X+iP_X}{\sqrt{2}},\quad \mathbf{y}=\frac{-Y+iP_Y}{\sqrt{2}},\label{eq:copx1}\\&&
\mathbf{p_x}=\frac{iX+P_X}{\sqrt{2}}, \quad \mathbf{p_y}=\frac{iY-P_Y}{\sqrt{2}}.\label{eq:copx2}
\end{eqnarray}Thus, if we express the Hamiltonian (\ref{eq:nh2f}) in eigen 
coordinates (\ref{eq:copx1}-\ref{eq:copx2}), we obtain 
normal form of $H_2$ i.e.
\begin{eqnarray}
 &&H_2=i\omega_1 XP_X-i\omega_2YP_Y.\label{eq:nmh2f}
\end{eqnarray} The above complexification gives  the final change used in this 
article as follows
\begin{eqnarray}
 &&\mathbf{C}=[c_{ij}], \, 1\leq i, j\leq 4,\label{eq:Ch2f}
\end{eqnarray}where
\begin{eqnarray*}
&&c_{11}=0;c_{12}=0;c_{13}=\frac{\mathbf{n}^2+Q+\omega _1{}^2}{\surd d(\omega_1)};\\&&c_{14}=\frac{\mathbf{n}^2+Q+\omega _2{}^2}{\surd d(\omega_2)};
c_{21}=\frac{2\mathbf{n} \omega _1}{\surd d(\omega_1)};\\&&c_{22}=\frac{2\mathbf{n} \omega_2}{\surd d(\omega_2)};c_{23}=\frac{-S}{\surd d(\omega_1)};\\&& c_{24}=\frac{-S}{\surd d(\omega_2)};
c_{31}=\frac{(-\mathbf{n}^2+Q) \omega _1+\omega _1{}^3}{\surd d(\omega_1)};\\&&c_{32}=\frac{(-\mathbf{n}^2+Q) \omega _2+\omega _2{}^3}{\surd d(\omega_2)};
c_{33}=\frac{\mathbf{n} S}{\surd d(\omega_1)};\\&&
c_{34}=\frac{\mathbf{n} S}{\surd d(\omega_2)};
c_{41}=\frac{- S \omega _1}{\surd d(\omega_1)};
c_{42}=\frac{- S \omega _2}{\surd d(\omega_2)};\\&&
c_{43}=\frac{\mathbf{n}^3+\mathbf{n} Q -\mathbf{n} \omega _1{}^2}{\surd d(\omega_1)};
c_{44}=\frac{\mathbf{n}^3+\mathbf{n} Q -\mathbf{n} \omega _2{}^2}{\surd d(\omega_2)},
\end{eqnarray*}
where $d(\omega_{1,2})$ are obtain from equation (\ref{eq:dw}). If we ignore the perturbations $A_2, \, \epsilon_1$ and $\epsilon_2$ then above symplectic change agree with that of \cite{Deprit1967AJ.....72..173D} and \cite{Mayer}.

\section{Nonlinear Stability of Triangular Point in Non-resonance Case} 
\label{sec:nstb}
To study the nonlinear stability of the equilibrium points, there are 
two cases: (i) resonance case and (ii) non-resonance case. Nonlinear stability in resonance case would be discussed with the help of theorems of Sokolsky (1974), Markeev (1978) and Grebenikov (1986) as in  \cite{K.Gozdzieski1998CeMDA..70...41G} whereas, in later case it is studied with the help of Arnold-Moser theorem. Here, authors are interested only to examine the nonlinear stability in  non-resonance case by the use of Arnold-Moser theorem under the influence of perturbations. The general form of the theorem is presented in Appendix (\ref{subsec:AM}.) as in \cite{Mayer,Hall1992ihds.book.....M} however, for self sufficient paper, authors have described the theorem in their own notations as follows:

Consider the Hamiltonian expressed in action-angle variables $(I_1,\,I_2,\,\phi_1,\,\phi_2$) as
\begin{eqnarray}
 K=K_2+K_4+K_6+\dots+K_{2n}+K^{*}_{2n+1},\label{eq:amk1}
\end{eqnarray}where $K_{2r}$ are homogeneous polynomials in action variables $I_1,\,I_2$, of degree $r$,  $K_{2n+1}$ are polynomials containing terms of higher order than $n$ and $K_2=\omega_1I_1-\omega_2I_2$ with $\omega_i,\, i=1,\,2$ positive constants. $K_4=-\left(K_{2020}I_1^{2}+K_{1111}I_1I_2+K_{0202}I_2^{2}\right)$, where $K_{2020}\,K_{1111},\,K_{0202}$ are constants. 
 Since, $K_2,\,K_4,\,\dots,\, K_{2n}$ are function of only action variables $I_1,\,I_2$, so, the Hamiltonian is assumed to be in Birkhoff's normal form up to terms of degree $2n$ which can be obtained with few non-resonance assumptions on the frequencies  $\omega_1$ and $\omega_2$, but in order to state the theorem, assume that $K$ is in the required form. Now, Arnold-Moser theorem stated as:
\begin{kam}[Arnold-Moser]
 The origin is stable for the system whose Hamiltonian is (\ref{eq:amk1}) provided for some $r,\,1\leq r\leq n,\, D_{2r}=K_{2r}(\omega_2,\,\omega_1)\neq 0$ or equivalently provided $K_2$ does not divide $K_{2r}.$ 
\end{kam}
 Since, Arnold-Moser theorem requires Birkhoff's normal form of the Hamiltonian and Birkhoff's normal form requires some non-resonance assumptions on the frequencies $\omega_1$ and $\omega_2$, which is described as follows \citep{Deprit1967AJ.....72..173D}. Suppose, $\omega_1$ and $\omega_2$ are frequencies in case of linear dynamics of the 
infinitesimal mass, and $n$ is an integer such that $n\geq 
2$, then  \begin{eqnarray}
l_1\omega_1+l_2\omega_2\neq 0,\label{eq:nrcnd}
\end{eqnarray} for all $l_1, \, l_2 \in\mathbb{Z}$ such that $|l_1|+|l_2|\leq 2n$ 
known as irrationality condition. This condition ensures that there 
is an analytic  symplectic normalizing transformation such that the Hamiltonian (\ref{eq:tehm1}) takes the form (\ref{eq:amk1}). Coefficients in the normalized Hamiltonian are 
independent on the integer $n$ and also independent to the manner of 
transformation is obtained. In particular, the determinant
\begin{eqnarray}\mbox{det}\begin{bmatrix}
\frac{\partial^{2} K}{\partial I^2_{1}}&\frac{\partial^{2} K}{\partial I_{1}\partial I_{2}}&\frac{\partial K}{\partial I_1} \\
\frac{\partial^{2} K}{\partial I_{2}\partial I_{1}}&\frac{\partial^{2} K}{\partial I^2_{2}}&\frac{\partial K}{\partial I_2}\\
\frac{\partial K}{\partial I_1}&\frac{\partial K}{\partial I_2}&0
\end{bmatrix}_{I_1, \, I_2=0}
\label{eq:kam}\end{eqnarray}
is an invariant of the Hamiltonian (\ref{eq:bhnorm}) with respect to the 
symplectic transformation used. Arnold-Moser theorem decides the 
stability of equilibrium points under these two conditions.

Here, we are interested to implement this procedure to the problem in question for $n=2$. That is, we have to compute Birkhoff's normal form of Hamiltonian (\ref{eq:tehm1}) up to degree $2$ in action variables and then analyze the quantity $D_4$ with respect to the perturbations in question.

\subsection{Birkhoff's Normal Form}
\label{subsec:norm2}
 In order to apply Arnold-Moser theorem, we have to compute Brikhoff's normal form up to $4$th order normal form of the
Hamiltonian in the vicinity of equilibrium point which 
will be the function of action-angle variables $(I_1, \, I_2, \, \phi_1, \, 
\phi_2)$. To obtain the Brikhoff's normal form, we have used Lie transform described in \cite{coppola1989ZaMM...69..275C}  and \cite{Jorba}. Since, higher order normalized Hamiltonian  is given as \citep{coppola1989ZaMM...69..275C}:
\begin{eqnarray}
&&K=K_2+K_3+K_4\dots,\label{eq:bhnorm} \end{eqnarray}
where 
\begin{eqnarray}
&&K_n=\sum{K_{jkls}}X^{j}Y^{k}P^{l}_XP^{s}_Y\quad \text{with}\nonumber\\&& j+k+l+s=n
\end{eqnarray} is known as Kamiltonian.  Now, the quadratic part of $n$-th order normal form $K_n$ is $K_2=H_2$.
At the $n^{th}$ step of the Lie transform method, Kamiltonian $K_n$ is obtained by the expression 
\begin{eqnarray}
K_n=\frac{1}{n}\left\{H_2, \, W_n\right\}+\text{previously known terms}.\label{eq:kn}
\end{eqnarray}
 Now,  to determine the generating function $W_n$, which provides the best simplified form of Kamiltonian $K_n$, first we determine the Lie bracket $\left\{H_2, \, W_n\right\}$ with $H_2$ from equation (\ref{eq:nmh2f}) as follows  
\begin{eqnarray}
  \left\{H_2, \, W_n\right\}&=&\frac{\partial H_2}{\partial X}\frac{\partial W_n}{\partial P_X}+\frac{\partial H_2}{\partial Y}\frac{\partial W_n}{\partial P_Y}\nonumber\\&&-\frac{\partial H_2}{\partial P_X}\frac{\partial W_n}{\partial X}-\frac{\partial H_2}{\partial P_Y}\frac{\partial W_n}{\partial Y}\nonumber\\
&=&i\omega_1\left[P_X\frac{\partial W_n}{\partial P_X}-X\frac{\partial W_n}{\partial X}\right]\nonumber\\&&-i\omega_2\left[P_Y\frac{\partial W_n}{\partial P_Y}-Y\frac{\partial W_n}{\partial Y}\right].\label{eq:wn}
\end{eqnarray}
We need $W_n$ such that results of this partial linear differential operator on $W_n$ remove out as many terms as possible in the expression of Kamiltonian $K_n$. 
 It is clear from the expression of $K_n$ that each term to be canceled  will be of the form $P_0X^{j}Y^{k}P_X^{l}P_Y^{s}$, where $P_0$ is constant. Let us take, $W_n$ as the sum of terms of the form $Q_0X^{j}Y^{k}P_X^{l}P_Y^{s}$, where $Q_0$ is an undetermined constant and hence, we get
\begin{eqnarray}
\frac{\left\{H_2, \, W_n\right\}}{n}=\frac{i\left[\omega_1(l-j)-\omega_2(s-k)\right]}{n}Q_0X^{j}Y^{k}P_X^{l}P_Y^{s},\label{eq:kn0}\end{eqnarray}
which gives
\begin{eqnarray}
Q_0&=&\frac{inP_0}{\omega_1(l-j)-\omega_2(s-k)}\quad \text{with}\nonumber\\&&j+k+l+s=n.\label{eq:Q0}\end{eqnarray}
Now, from equation (\ref{eq:Q0}), it is clear that above scheme fails if denominator vanishes. If we assume that frequencies $\omega_{1,2}$ are non-resonant then denominator will vanish only in case of $l=j$ and $s=k$. In other words, in the expression of $K_n$, the terms of the type $(X^{l}Y^{s}P_X^{l}P_Y^{s})$ can not be removed. Hence, in case of non-resonance, the problem always can be reduced to the form of (\ref{eq:bhnorm}), where
\begin{eqnarray}
K_2&=&H_2=i\omega_1 XP_X-i\omega_2YP_Y,\\
K_3&=&0,\\
K_4&=&K_{2020}X^{2}P_X^{2}+K_{1111}XYP_XP_Y\nonumber\\&&+K_{0202}Y^{2}P_Y^{2}.\label{eq:KI1}
\end{eqnarray}
Thus, in case of non-resonance, normalized (transformed) Hamiltonian is a function of only action variables, $I_1=iXP_X$ and $I_2=iYP_Y$ and hence, both the coordinates are ignorable which infer that system is consistent and system is said to be in Brikhoff's normal form. Now, in term of action variable $I_1, \, I_2$, $4$th order part of normalized Hamiltonian is written as
\begin{eqnarray}
K_4(I_1, \, I_2)&=&-\left(K_{2020}I_1^{2}+K_{1111}I_1I_2\right.\nonumber\\&&\left.+K_{0202}I_2^{2}\right).\label{eq:KI2}\end{eqnarray}
On the other hand, in case of resonant values of frequencies $\omega_{1,2}$, some additional non-removable terms occur while solving the generating function $W_n$. For simplicity, coefficients $K_{jkls}$ in equation (\ref{eq:KI2}) are split-up into four parts such as $\mathbf{k}_{jkls}, \, \mathbf{k}_{jklse1}$,  $\mathbf{k}_{jklsA}$ and $\mathbf{k}_{jklse2}, \, j, \, k, \, l, \, s=0, \, 1, \, 2,\,3,\,4$ such that $j+k+l+s=4$. These coefficients correspond to the terms due to classical case,  radiation effects $\epsilon_1$, oblateness $A_2$ and presence of the disc $\epsilon_2$, respectively. In other words

\begin{eqnarray}
K_{2020}&=& \mathbf{k}_{2020}+\mathbf{k}_{2020e1}+\mathbf{k}_{2020A}\nonumber\\&&+\mathbf{k}_{2020e2},\label{eq:k20}\end{eqnarray}
\begin{eqnarray}
K_{1111}&=& \mathbf{k}_{1111}+\mathbf{k}_{1111e1}+\mathbf{k}_{1111A}\nonumber\\&&+\mathbf{k}_{1111e2},\label{eq:k11}\end{eqnarray}
\begin{eqnarray}
K_{0202}&=& \mathbf{k}_{0202}+\mathbf{k}_{0202e1}+\mathbf{k}_{0202A}\nonumber\\&&+\mathbf{k}_{0202e2},\label{eq:k02}\end{eqnarray}
where $\mathbf{k}_{jkls}$ represents coefficients for classical part, $\mathbf{k}_{jklse1}$ indicates coefficients for radiation pressure terms, $\mathbf{k}_{jklsA}$ used for oblateness and $\mathbf{k}_{jklse2}$ for the disc with $j, \, k, \, l, \, s=0, \, 1, \, 2,\,3,\,4$ such that $j+k+l+s=4$. But, in the absence of perturbations i.e. $A_2=\epsilon_1=\epsilon_2=0$, $K_{jkls}=\mathbf{k}_{jkls}$, which are equivalent to the coefficients of the normalized Hamiltonian of the classical problem. Further, due to very large expressions of all the coefficients $\mathbf{k}_{jkls}$, $\mathbf{k}_{jklse1}$, $\mathbf{k}_{jklsA}$ and  $\mathbf{k}_{jklse2}$ with $j, \, k, \, l, \, s=0, \, 1, \, 2,\,3,\,4$  such that $j+k+l+s=4$, in equations  (\ref{eq:k20}-\ref{eq:k02}), these are given in Appendix \ref{subsec:coefK2}.

As, we have the fourth order normalized Hamiltonian given in 
equation (\ref{eq:KI2}), so, we have computed the determinant $D_4$ to use Arnold-Moser theorem which will decide 
 nonlinear stability of the triangular equilibrium point in non-resonance case. In our case, we found $D_4$ as:
\begin{eqnarray}
&&D_4=D_{40}+D_{41}\epsilon_1+D_{{42}} A_2+D_{43}\epsilon_2
, \label{eq:d2} 
\end{eqnarray}
where $D_{40}, \, D_{41}, \, D_{42}$ and $D_{43}$  correspond to the classical parts, radiation pressure term, oblateness and the terms due to presence of the disc, respectively, which are obtained with the help of equations (\ref{eq:KI1},\ref{eq:k20}-\ref{eq:k02}) and have very large expression hence, these are given in Appendix \ref{subsec:coefD2}. 
Substituting the values of $D_{40}, \, D_{41}, \, D_{42}$ and $D_{43}$ in terms of frequencies $\omega_{1,2}$ from Appendix \ref{subsec:coefD2} into equation (\ref{eq:d2}) and after a tedious simplification at several stages, we get
\begin{eqnarray}
 D_4&=&\frac{-36+541 \omega _1^2 \omega _2^2-644 \omega _1^4 \omega _2^4}{8 \left(1-4\omega_1^2 \omega _2^2\right) \left(-4+25 \omega _1^2 \omega _2^2\right)}+\frac{G_1 \epsilon _1}{G_2}\nonumber\\&&+\frac{G_3 A_2}{G_4}+\frac{G_5 \epsilon _2}{G_6},\label{eq:D2f}
\end{eqnarray}
where\begin{eqnarray}
G_1&=&\left(691394292-5589532701 \omega _1^2 \omega _2^2\right.\nonumber\\&&\left.+7636835570 \omega _1^4 \omega _2^4-4919234984 \omega _1^6 \omega _2^6\right.\nonumber\\&&\left.+373115136 \omega _1^8 \omega _2^8\right.\nonumber\\&&\left.+597854208 \omega _1^{10} \omega _2^{10}\right),\end{eqnarray}
\begin{eqnarray}
G_2&=&12288 \omega _1^2 \omega _2^2 \left[\left(1-4 \omega _1^2 \omega _2^2\right){}^2\left(4-25 \omega _1^2 \omega _2^2\right)\right.\nonumber\\&&\left.\left(117+16 \omega _1^2 \omega _2^2\right)\right],\end{eqnarray}
\begin{eqnarray}
G_3&=&\left(103351788-881740119 \omega _1^2 \omega _2^2\right.\nonumber\\&&\left.+1442778388 \omega _1^4 \omega _2^4-916084992 \omega _1^6 \omega _2^6\right.\nonumber\\&&\left.+193442816 \omega _1^8 \omega _2^8\right),\end{eqnarray}
\begin{eqnarray}
G_4&=&384 \omega _1 \omega _2 \left[\left(1-4 \omega _1^2 \omega _2^2\right){}^2\left(-4+25 \omega _1^2 \omega _2^2\right)\right.\nonumber\\&&\left. \left(117+16 \omega _1^2 \omega _2^2\right)\right],\end{eqnarray}
\begin{eqnarray}
G_5&=&\left(341649252-2463280353 \omega _1^2 \omega _2^2\right.\nonumber\\&&\left.+1827505940 \omega _1^4 \omega _2^4-1159472852 \omega _1^6 \omega _2^6\right.\nonumber\\&&\left.+210281632 \omega _1^8 \omega _2^8\right.\nonumber\\&&\left.+205320704 \omega _1^{10} \omega _2^{10}\right),\end{eqnarray}
\begin{eqnarray}
G_6&=&1536 \omega _1^2 \omega _2^2 \left(1-4 \omega _1^2 \omega _2^2\right){}^2 \left(-8910\right.\nonumber\\&&\left.+2861 \omega _1^2 \omega _2^2+400 \omega _1^4 \omega _2^4\right).
\end{eqnarray}

\begin{figure}
\includegraphics[scale=.65]{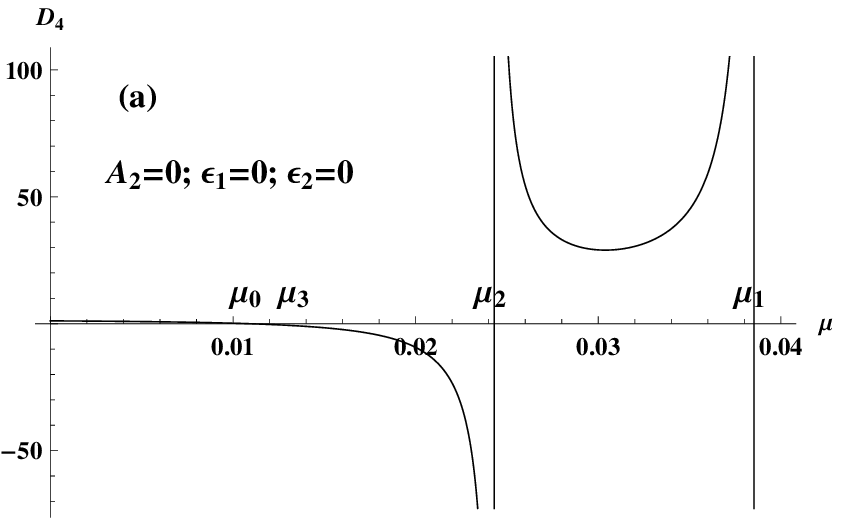}\\\includegraphics[scale=.65]{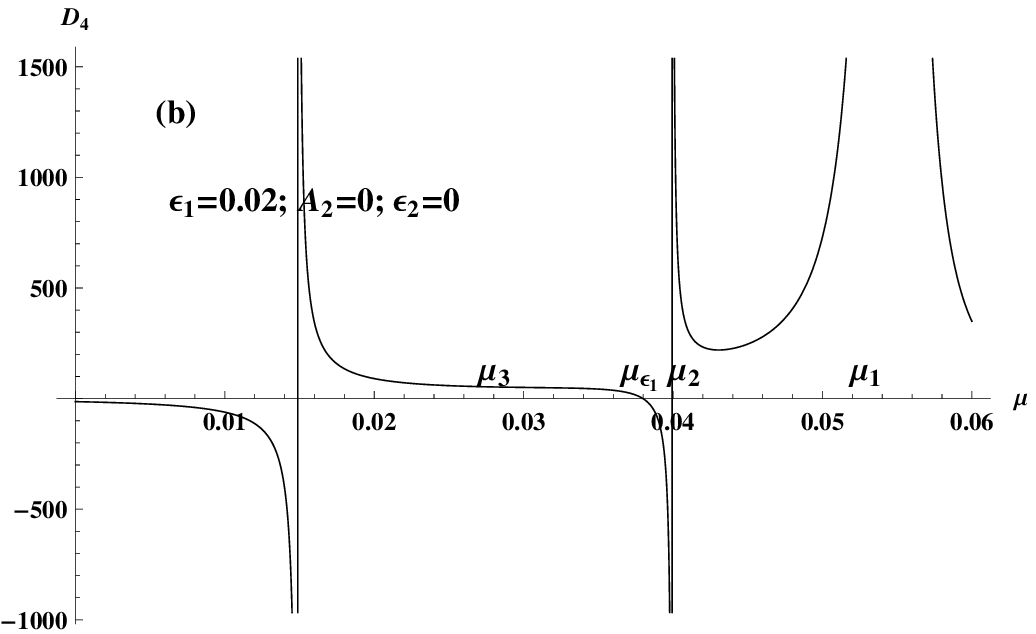}\\\includegraphics[scale=.65]{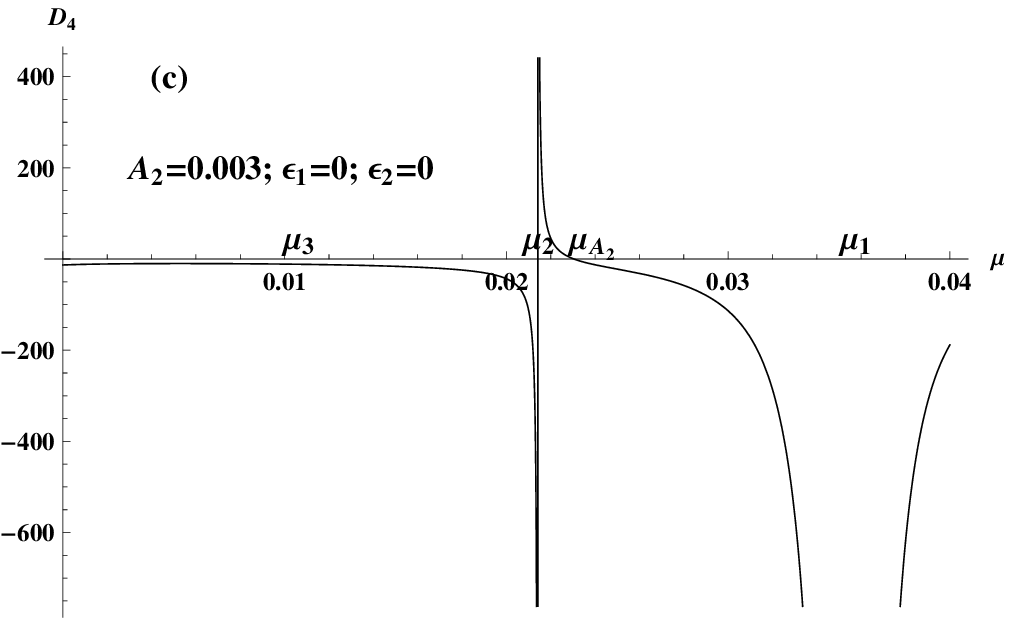}\\\includegraphics[scale=.65]{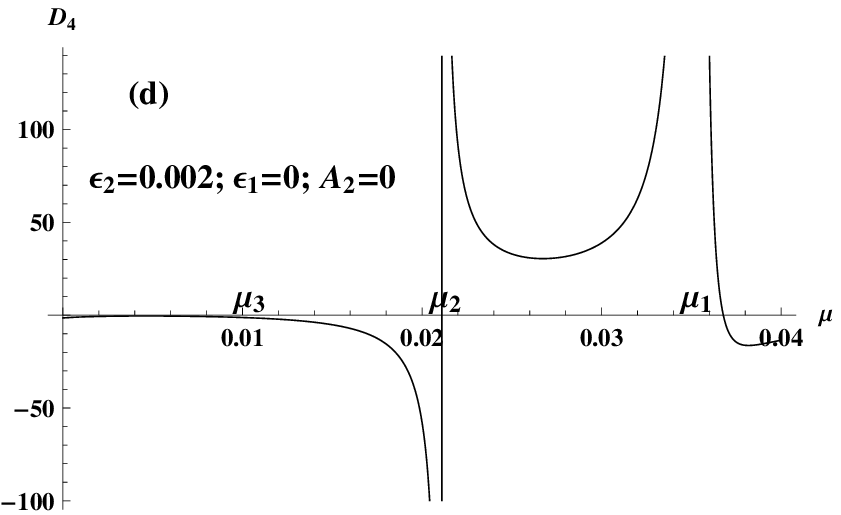}\\\includegraphics[scale=.65]{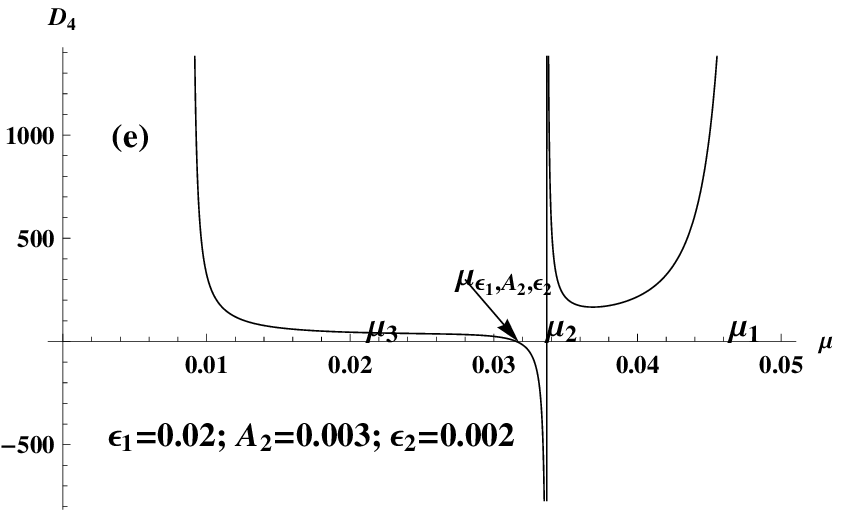}\\\includegraphics[scale=.55]{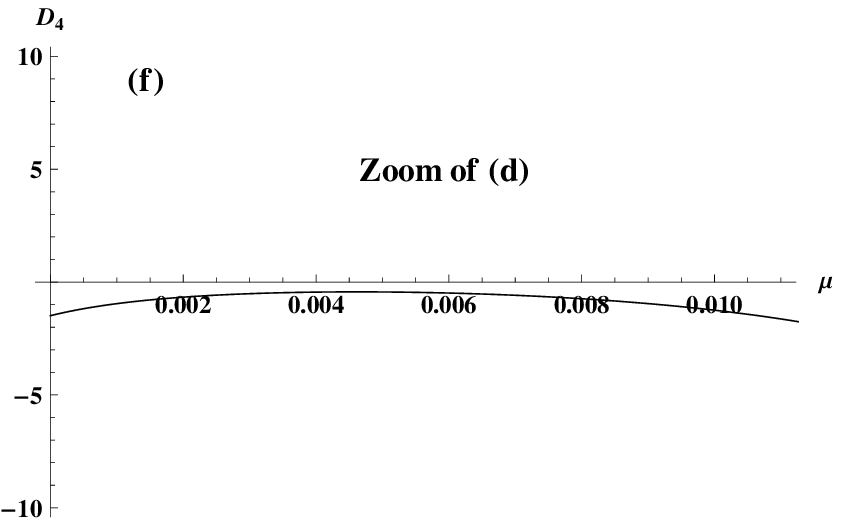}      
\caption{Stability condition from the normalized Hamiltonian of order four.\label{fig:dstb1}}
\end{figure}
\begin{figure}
\includegraphics[scale=.65]{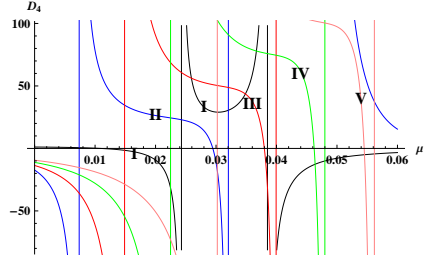}
\caption{Change in critical value $\mu_{\epsilon_1}$ with respect to radiation pressure parameter $\epsilon_1=1-q_1$. Curves are drawn at $A_2=0.0,\, \epsilon_2=0.0$ and I: $\epsilon_1=0.0,$ II :$\epsilon_1=0.01,$ III: $\epsilon_1=0.02,$  IV: $\epsilon_1=0.03,$ V: $\epsilon_1=0.04.$\label{fig:dstb2}}
\end{figure}
\begin{figure}
\includegraphics[scale=.65]{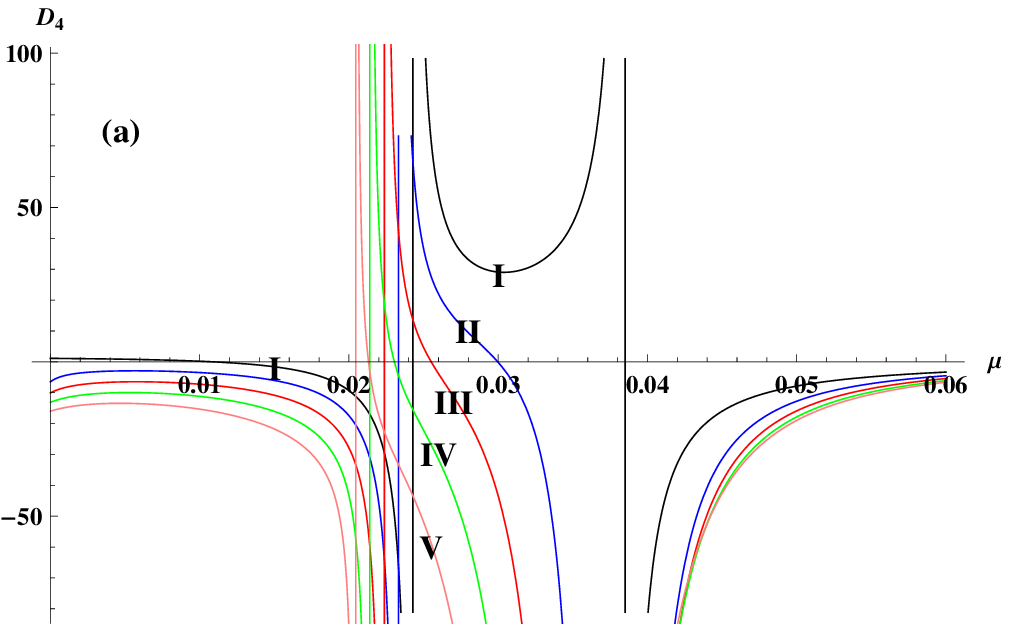}\\\includegraphics[scale=.65]{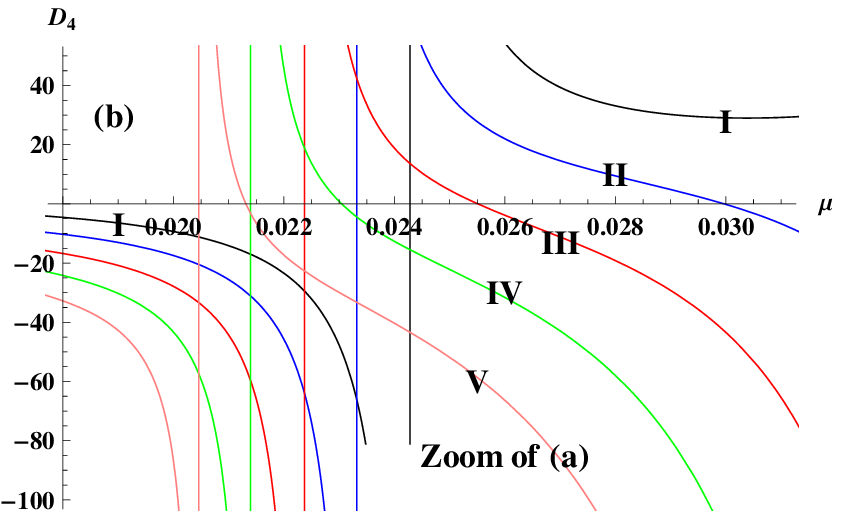}
\caption{Change in critical value $\mu_{A_2}$ with respect to oblateness parameter $A_2$. (a) Curves are drawn at $\epsilon_1=0.0,\, \epsilon_2=0.0$ and I: $A_2=0.0,$ II :$A_2=0.001,$ III: $A_2=0.002,$  IV: $A_2=0.003,$ V: $A_2=0.004.$ (b) Zoom of figure (a).\label{fig:dstb3}}
\end{figure}
\begin{figure}
\includegraphics[scale=.65]{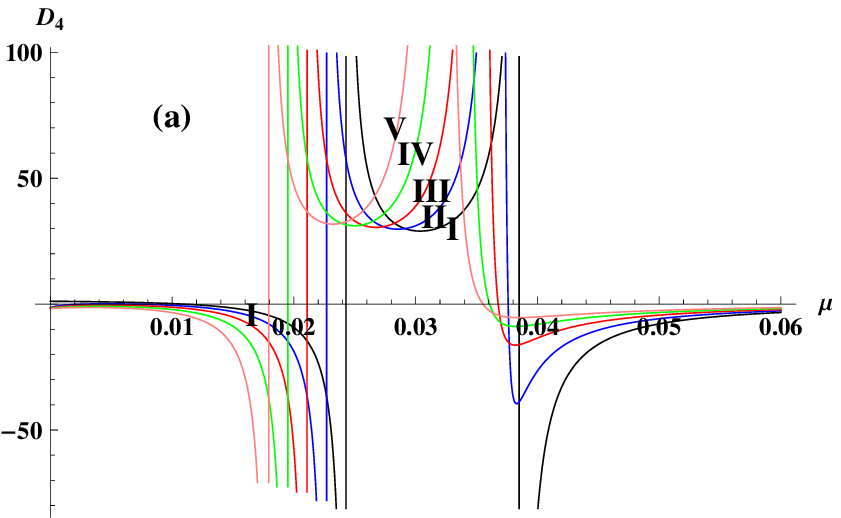}\\\includegraphics[scale=.65]{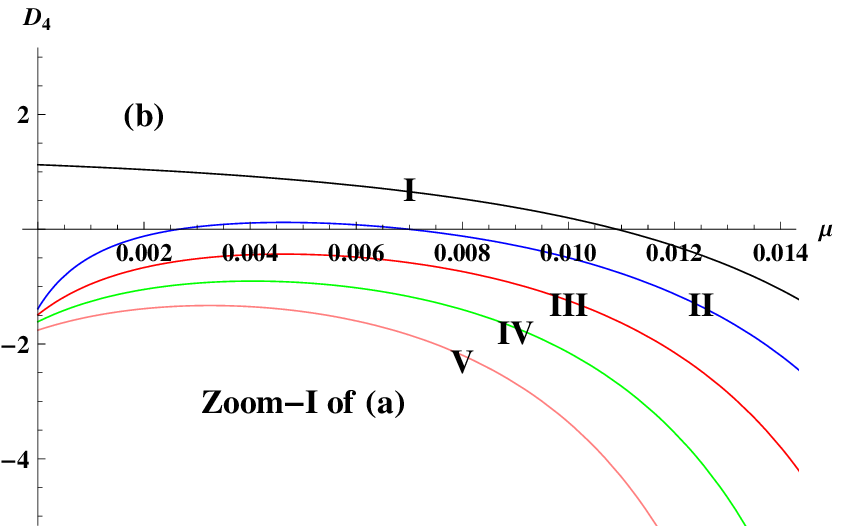}\\\includegraphics[scale=.65]{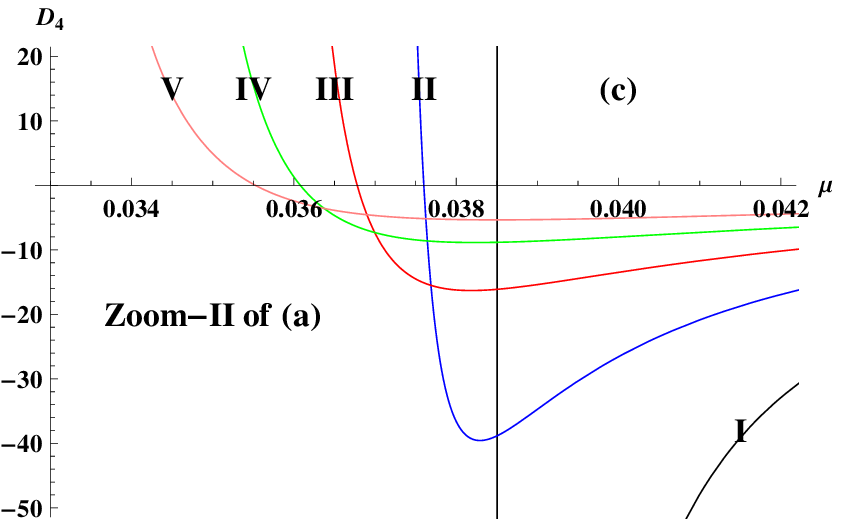}
\caption{Change in critical value $\mu_{\epsilon_2}$ with respect to disc parameter $\epsilon_2=b-1$. (a) Curves are drawn at $\epsilon_1=0.0,\, A_2=0.0$ and I: $\epsilon_2=0.0,$ II :$\epsilon_2=0.001,$ III: $\epsilon_2=0.002,$  IV: $\epsilon_2=0.003,$ V: $\epsilon_2=0.004.$ (b) Zoom-I of figure (a). (c) Zoom-II of figure (a).\label{fig:dstb4}}
\end{figure}

\begin{deluxetable}{rrrrr}
\tabletypesize{\scriptsize}
\tablecaption{Stability range and critical values of $\,\mu\,$  at which $\,D_4=0\,$ for different values of perturbation parameters.\label{tbl-1}}
\tablewidth{0pt}
\tablehead{
\colhead{$\epsilon_1$}& \colhead{$A_2$} & \colhead{$\epsilon_2$} & \colhead{$\text{Value of}\,\mu\, \text{at which}\, D_4=0$} & \colhead{$\text{stability range}\,0<\mu<\mu_1=\bar{\mu_c}$} 
}
\startdata
0.0 & 0.0 & 0.0&0.01095&$(0.0,\, 0.03852)$\\
0.01 & 0.0 & 0.0&0.02949&$(0.0,\, 0.03861)$\\
0.02 & 0.0 & 0.0&0.03799&$(0.0,\, 0.03870)$\\
0.03 & 0.0 & 0.0&0.04626&$(0.0,\, 0.03879)$\\
0.04 & 0.0 & 0.0&0.05447&$(0.0,\, 0.03888)$\\
&&&&\\
0.0 & 0.001 & 0.0&0.02999&$(0.0,\, 0.03838)$\\
0.0 & 0.002 & 0.0&0.02553&$(0.0,\, 0.03825)$\\
0.0 & 0.003 & 0.0&0.02305&$(0.0,\, 0.03811)$\\
0.0 & 0.004 & 0.0&0.02132&$(0.0,\, 0.03798)$\\
&&&&\\
0.0 & 0.0 & 0.001&0.00269&$(0.0,\, 0.03853)$\\
0.0 & 0.0 & 0.001&0.00697&$(0.0,\, 0.03853)$\\
0.0 & 0.0 & 0.001&0.03761&$(0.0,\, 0.03853)$\\
0.0 & 0.0 & 0.002&0.03678&$(0.0,\, 0.03854)$\\
0.0 & 0.0 & 0.003&0.03608&$(0.0,\, 0.03855)$\\
0.0 & 0.0 & 0.004&0.03551&$(0.0,\, 0.03855)$\\
0.0 & 0.0 & 0.007&0.03506&$(0.0,\, 0.03858)$\\
0.0 & 0.0 & 0.009&0.03706&$(0.0,\, 0.03859)$\\
0.0 & 0.0 & 0.010&0.03925&$(0.0,\, 0.03860)$\\
0.0 & 0.0 & 0.030&0.05261&$(0.0,\, 0.03877)$\\
0.0 & 0.0 & 0.060&0.07054&$(0.0,\, 0.03901)$\\
0.0 & 0.0 & 0.090&0.05604&$(0.0,\, 0.03926)$\\
0.0 & 0.0 & 0.100&0.05757&$(0.0,\, 0.03934)$\\
0.0 & 0.0 & 0.200&---&$(0.0,\, 0.04015)$\\
0.0 & 0.0 & 0.300&0.08642&$(0.0,\, 0.04097)$\\
&&&&\\
0.02 & 0.003 & 0.0&0.03486&$(0.0,\, 0.03830)$\\
0.02 & 0.0 & 0.002&0.03474&$(0.0,\, 0.03873)$\\
0.0 & 0.003 & 0.002&0.01978&$(0.0,\, 0.03813)$\\
0.02 & 0.003 & 0.002&0.03155&$(0.0,\, 0.03833)$\\
\enddata
\end{deluxetable}

From the expression (\ref{eq:D2f}) of $D_4$, it can be easily seen that if 
perturbations are ignored then it coincides with that of 
\cite{Deprit1967AJ.....72..173D,coppola1989ZaMM...69..275C,Shevchenko2008CoPhC.178..665S} and \cite{Alvarez-Ram2012arXiv1212.2179A}.  To see the effect of perturbations on the nonlinear stability, we plot 
the expression of $D_4$ against mass ratio $\mu$ at different values of 
perturbations, individually as well as simultaneously (Fig. \ref{fig:dstb1}-\ref{fig:dstb4}), in 
the stability region $0<\mu<\mu_1=\bar{\mu_c}$. 
From  figure (\ref{fig:dstb1}a), it is seen that there 
is only one zero of $D_4$ at $\mu_0=0.01095$ whereas, $\mu_\kappa, \, \kappa=1,2,3$  
indicate the values of mass ratio under three main resonance cases of linear stability \citep{Kishor2013MNRAS.436.1741K}, which change their values with respect to the values of $\epsilon_1, A_2, \epsilon_2$ (Fig. \ref{fig:dstb1} a-e). Thus, Arnold-Moser theorem infer that the triangular equilibrium point $L_4$  is stable in the interval  $0<\mu<\mu_1$ except at $\mu_0=0.01095$, in addition to  the values $\mu_\kappa, \, \kappa=1,2,3$ for three main 
resonance cases of linear stability. This result is similar 
to that of \cite{Deprit1967AJ.....72..173D} in the absence of perturbations. 
The presence of radiation pressure (Fig. \ref{fig:dstb1} b) as well as  oblateness (Fig. \ref{fig:dstb1} c) cause the theorem fail at $\mu_{\epsilon_1}=0.03795$ and  $\mu_{A_2}=0.02297$, respectively. The effect of disc seen only at very small values of 
$\epsilon_2$ (Fig. \ref{fig:dstb1} d) and the value of $\mu$ at which $D_4$ vanishes is $0.03678$ (Fig. \ref{fig:dstb1} d) whereas, at larger value of $\epsilon_2$, it is beyond the 
stability range $0<\mu<\mu_1$.  If, we analyze, by taking two perturbations at a time such as $(\epsilon_1, \, A_2), \, (\epsilon_1, \, \epsilon_2)$ and $(A_2, \, \epsilon_2)$ then zeros of $D_4$ i.e. the value of mass ratio $\mu$ at which  $D_4$ vanish, are obtained as $0.03486, \, 0.03474$ and $0.01978$, respectively. If all the perturbations taken at a time then 
the value of the mass ratio is $\mu_{\epsilon_1, A_2, \epsilon_2}=0.03155$ at which theorem fails.  Thus, the triangular equilibrium point within the stability range, in the absence as well as presence of perturbation parameters,  is unstable due to failure of Arnold-Moser theorem. In order to see the effect of perturbation, we have drawn the figures (\ref{fig:dstb2}-\ref{fig:dstb4}) by varying individual parameter taking remaining two of them zero. From, figure (\ref{fig:dstb2}), critical value $\mu_{\epsilon_1}$ increases with the increment in the value of $\epsilon_1$. The critical values $\mu_{\epsilon_1}$ corresponding to the curve I, II and III lie within the stability range whereas for curve IV and V, it is beyond the stability range (see Table-\ref{tbl-1}). From figure (\ref{fig:dstb3}a,b) and Table-\ref{tbl-1}, it is clear that the critical value $\mu_{A_2}$ decreasing with the increment in the value of $A_2$ (also, see Table-\ref{tbl-1}). Figure (\ref{fig:dstb4}a,b) shows that there are three critical value $\mu_{\epsilon_2}$ at $\epsilon_2=0.001$ but for other values it reduces to one.  It is also, clear that value of $\mu_{\epsilon_2}$ decreasing slowly with increment in the value of $\epsilon_2$, but when, value of $\epsilon_2$ increases after $0.009$, the value of critical $\mu_{\epsilon_2}$ go beyond the corresponding stability range (Table-\ref{tbl-1}).

\section{Conclusion}
\label{sec:con}
The analysis of nonlinear stability of triangular equilibrium point in the Chermnykh-like problem under the influence of perturbations have been preformed for  non-resonance case. The diagonalization and higher order normalization of Hamiltonian of the problem are made by the method of Lie transform under the influence of perturbations in the form of radiation pressure force, oblateness and the disc. Due to perturbations, transformation equations take a complicated form but in the absence of perturbations these equations agree with the classical results. Next, we have analyzed nonlinear stability with the help of Arnold-Moser theorem. After a huge and tedious computation at several intermediate steps, we have obtained determinant $D_4$ in the presence of perturbations which  is agree with  that of  \cite{Deprit1967AJ.....72..173D, Mayer, coppola1989ZaMM...69..275C} and \cite{Shevchenko2008CoPhC.178..665S} under the vanishing condition of perturbations. Due to Arnold-Moser theorem, it is found that under the influence of perturbations, the motion of infinitesimal mass in the vicinity of triangular equilibrium point get affected. In other words, in classical case, triangular equilibrium point is stable within the stability range  $0<\mu<\mu_1=0.03852$ except for the value of $\mu_2, \, \mu_3$ and $\mu_0=0.01095$ at which $D_4$ vanishes and hence, Arnold-Moser theorem fails there. But under the effect of radiation pressure and oblateness, nonlinear stability of the triangular equilibrium point fail at $\mu_{\epsilon_1}=0.03795$ and $\mu_{A_2}=0.02297$ due to the same reason whereas, in the presence of all three perturbations, it fails at $\mu_{\epsilon_1,A_2,\epsilon_2}=0.03155$. The effect of disc is seen either at very small values of disc outer radius or very large values of disc outer radius. It is observed that for small values of disc outer radius, theorem fails at  $\mu_{\epsilon_2}=0.03678$, whereas for large value of disc outer radius, effect is beyond the stability range $0<\mu<\mu_1$. Moreover, in case of taking two perturbations at a time such as $(\epsilon_1, \, A_2), \, (\epsilon_1, \, \epsilon_2)$ and $(A_2, \, \epsilon_2)$, the zeros of $D_4$ are $0.03486, \, 0.03474$ and $0.01978$, respectively.  The nature of variation in the value of critical mass $\mu$ at which the value of $D_4=0$ at different values of individual perturbation parameters is observed and significant variation are found. The critical value $\mu_{\epsilon_1}$ first, increases with the increment in the value of $\epsilon_1$, and then this value go beyond the stability range (see figure \ref{fig:dstb2} and Table-\ref{tbl-1}). Again, the critical value $\mu_{A_2}$ decreasing with the increment in the value of $A_2$ (see figure \ref{fig:dstb3}a,b and Table-\ref{tbl-1}). There are three critical value $\mu_{\epsilon_2}$ at $\epsilon_2=0.001$ Figure (\ref{fig:dstb4}a,b) but for other values it reduces to one.  It is also, clear that value of $\mu_{\epsilon_2}$ decreasing slowly with increment in the value of $\epsilon_2$, but for larger values of $\epsilon_2$, $\mu_{\epsilon_2}$ do not lie in the corresponding stability range (Table-\ref{tbl-1}). The results, which are obtained, are very helpful to observe the motion of infinitesimal mass such as spacecraft, asteroid or satellite  in the Sun-Jupiter system. The present study and observations are applicable to the analysis of more generalized problems and would be extended up to higher order in addition with some other type of perturbations like P-R drag, solar wind drag.  On the other hand results are limited up to radially symmetric disc but in future it would be extended. 
\begin{acknowledgments}
The authors are very thankful for the referee's comments and suggestions; they have been very useful and have greatly improved the manuscript.  The  financial support by the Department of Science and Technology, Govt. of India through the SERC-Fast Track Scheme for Young Scientist [SR/FTP
/PS-121/2009] is duly acknowledged. Some of the important references in addition to basic facilities are provided by IUCAA Library, Pune, India.\end{acknowledgments}

% \bibliographystyle{aps-nameyear}      % American Physical Society (APS) style, author-year citations
% \bibliography{ref} 
%% The Appendices part is started with the command \appendix;
%% appendix sections are then done as normal sections
%% \appendix
% \onecolumn
% \appendix
\appendix
% \twocolumn[appendix]
\section{Appendix}

\subsection{Arnold-Moser theorem \citep{Mayer,Hall1992ihds.book.....M}}
 \label{subsec:AM}

Consider a Hamiltonian, which is the function of canonical coordinates $x_i, \, y_i,\, i=1,\, 2,$ expressed as
\begin{eqnarray}
 H=H_2+H_4+H_6+\dots+H_{2n}+H^{*}_{2n+1},\label{eq:am1}
\end{eqnarray}where
\begin{enumerate}
 \item  $H$ is real analytic in the a neighborhood of the origin in $\mathbb{R}^4$;
\item  $H_{2k},\, 1\leq k\leq n,$ is a homogeneous function of degree $k$ in $I_i=\frac{1}{2}(x^{2}_i+y^{2}_i),\, i=1,\,2$;
\item $H^*$ has a series expansion which starts with terms at least of order $2n+1$;
\item  $H_2=\omega_1I_1-\omega_2I_2$ with $\omega_i,\, i=1,\,2$ positive constants;
\item  $H_4=\frac{1}{2}\left(AI^{2}_1-2BI_1I_2+CI^{2}_2\right),\, A,\,B,\,C$ constants. 
\end{enumerate}
There are several implicit assumptions in stating that Hamiltonian $H$ in the form of (\ref{eq:am1}). As, $H$ is at least quadratic in canonical coordinates  $x_i, \, y_i,\, i=1,\, 2,$ the origin is assumed to be the equilibrium point in question. Again, $H_2=\omega_1I_1-\omega_2I_2$ is the Hamiltonian of two harmonic oscillators with frequency $\omega_1$ and $\omega_2$, the linearization at the origin of the system of equations whose Hamiltonian is $H,$ is two harmonic oscillators. Since, $H_2$ is not sign definite, a simple appeal to stability theorem of Lyapunov can not be made. Again, $H_2,\,H_4,\,\dots,\, H_{2n}$ are function of only $I_i=\frac{1}{2}(x_i+y_i),\, i=1,\,2,$ the Hamiltonian is assumed to be in Birkhoff's normal form up to terms of degree $2n$. The Birkhoff's normal form usually requires some non-resonance assumptions on the frequencies  $\omega_1$ and $\omega_2$, but in order to state the theorem, assume that $H$ is in the required form.

\begin{kam}[Arnold-Moser]
 The origin is stable for the system whose Hamiltonian is (\ref{eq:am1}) provided for some $k,\,2\leq k\leq n,\,\, D_{2k}=H_{2k}(\omega_2,\,\omega_1)\neq 0$ or equivalently provided $H_2$ does not divide $H_{2k}.$ 
\end{kam}

\subsection{Coefficient in $K_2$}
 \label{subsec:coefK2}
\begin{eqnarray}
{k_{2020}}&=&\frac{1}{\omega
_1 \left(2 \omega _1-\omega _2\right) \omega _2 \left(2 \omega _1+\omega _2\right)}\left[-4 i {g_{1011}} {g_{1110}} \omega _1^3-2 i {g_{0120}}{g_{2001}} \omega _1^2 \omega _2+\right.\nonumber\\&&\left.12 i {g_{1020}}{g_{2010}} \omega _1^2 \omega _2+2 i {g_{0021}} {g_{2100}} \omega _1^2 \omega _2+12 i {g_{0030}} {g_{3000}} \omega _1^2 \omega _2+\right.\nonumber\\&&\left.4 {g_{2020}}
\omega _1^3 \omega _2+i {g_{1011}} {g_{1110}} \omega _1 \omega _2^2+i{g_{0120}} {g_{2001}} \omega _1 \omega _2^2+i {g_{0021}}{g_{2100}}
\omega _1 \omega _2^2\right.\nonumber\\&&\left.-3 i {g_{1020}} {g_{2010}} \omega _2^3-3 i {g_{0030}}{g_{3000}} \omega _2^3-{g_{2020}} \omega _1 \omega _2^3\right],
\end{eqnarray}
\begin{eqnarray}
k_{2020e1}&=&
\frac{1}{\omega _1 \left(2 \omega _1-\omega _2\right) \omega _2 \left(2 \omega _1+\omega _2\right)}\left[-4 i {g_{1011e1}} {g_{1110}} \omega
_1^3-4 i {g_{1011}} {g_{1110e1}} \omega _1^3\right.\nonumber\\&&\left.-2 i {g_{0120e1}} {g_{2001}} \omega _1^2 \omega _2-2 i {g_{0120}} {g_{2001e1}} \omega _1^2
\omega _2+12 i {g_{1020e1}} {g_{2010}} \omega _1^2 \omega _2\right.\nonumber\\&&\left.+12 i {g_{1020}} {g_{2010e1}} \omega _1^2 \omega _2+2 i {g_{0021e1}} {g_{2100}}
\omega _1^2 \omega _2+2 i {g_{0021}} {g_{2100e1}} \omega _1^2 \omega _2\right.\nonumber\\&&\left.+
12 i {g_{0030e1}} {g_{3000}} \omega _1^2 \omega _2+12 i {g_{0030}}{g_{3000e1}} \omega _1^2 \omega _2+4 g_{2020e1} \omega _1^3 \omega
_2+\right.\nonumber\\&&\left.i {g_{1011e1}} {g_{1110}} \omega _1 \omega _2^2+i {g_{1011}} {g_{1110e1}} \omega _1 \omega _2^2+i {g_{0120e1}} {g_{2001}} \omega _1
\omega _2^2\right.\nonumber\\&&\left.+i {g_{0120}} {g_{2001e1}} \omega _1 \omega _2^2+i {g_{0021e1}} {g_{2100}} \omega _1 \omega _2^2+i {g_{0021}} {g_{2100e1}}
\omega _1 \omega _2^2\right.\nonumber\\&&\left.-3 i {g_{1020e1}} {g_{2010}} \omega _2^3-
3 i {g_{1020}} {g_{2010e1}} \omega _2^3-3 i {g_{0030e1}} {g_{3000}} \omega _2^3\right.\nonumber\\&&\left.-3 i {g_{0030}} {g_{3000e1}} \omega _2^3-{g_{2020e1}}
\omega _1 \omega _2^3\right],
\end{eqnarray}
\begin{eqnarray}
k_{2020A}&=&\frac{1}{\omega _1 \left(2 \omega _1-\omega _2\right) \omega _2 \left(2 \omega _1+\omega _2\right)}\left[-4 i {g_{1011A}} {g_{1110}} \omega
_1^3-4 i {g_{1011}} {g_{1110A}} \omega _1^3\right.\nonumber\\&&\left.-2 i {g_{0120A}} {g_{2001}} \omega _1^2 \omega _2-2 i{g_{0120}} {g_{2001A}} \omega _1^2 \omega
_2+12 i {g_{1020A}} {g_{2010}} \omega _1^2 \omega _2\right.\nonumber\\&&\left.+12 i {g_{1020}} {g_{2010A}} \omega _1^2 \omega _2+2 i {g_{0021A}} {g_{2100}} \omega
_1^2 \omega _2+2 i {g_{0021}} {g_{2100A}} \omega _1^2 \omega _2\right.\nonumber\\&&\left.+
12 i {g_{0030A}} {g_{3000}} \omega _1^2 \omega _2+12 i {g_{0030}} {g_{3000A}} \omega _1^2 \omega _2+4 {g_{2020A}} \omega _1^3 \omega
_2+\right.\nonumber\\&&\left.i {g_{1011A}} {g_{1110}} \omega _1 \omega _2^2+i {g_{1011}} {g_{1110A}} \omega _1 \omega _2^2+i {g_{0120A}} {g_{2001}} \omega _1 \omega
_2^2\right.\nonumber\\&&\left.+i {g_{0120}} {g_{2001A}} \omega _1 \omega _2^2+i {g_{0021A}} {g_{2100}} \omega _1 \omega _2^2+i {g_{0021}} {g_{2100A}} \omega _1
\omega _2^2\right.\nonumber\\&&\left.-3 i {g_{1020A}} {g_{2010}} \omega _2^3-
3 i {g_{1020}} {g_{2010A}} \omega _2^3-3 i {g_{0030A}} {g_{3000}} \omega _2^3\right.\nonumber\\&&\left.-3 i {g_{0030}} {g_{3000A}} \omega _2^3-{g_{2020A}}
\omega _1 \omega _2^3\right],
\end{eqnarray}
\begin{eqnarray}
k_{2020e2} &=&
\frac{1}{\omega _1 \left(2 \omega _1-\omega _2\right) \omega _2 \left(2 \omega _1+\omega _2\right)}\left[-4 i {g_{1011e2}} {g_{1110}} \omega
_1^3-4 i {g_{1011}} {g_{1110e2}} \omega _1^3\right.\nonumber\\&&\left.-2 i {g_{0120e2}} {g_{2001}} \omega _1^2 \omega _2-2 i {g_{0120}} {g_{2001e2}} \omega _1^2
\omega _2+12 i {g_{1020e2}} {g_{2010}} \omega _1^2 \omega _2\right.\nonumber\\&&\left.+12 i {{g1020}}{g_{2010e2}} \omega _1^2 \omega _2+2 i {g_{0021e2}} {g_{2100}}
\omega _1^2 \omega _2+2 i {g_{0021}} {g_{2100e2}} \omega _1^2 \omega _2\right.\nonumber\\&&\left.+
12 i {g_{0030e2}} {g_{3000}} \omega _1^2 \omega _2+12 i {g_{0030}} {g_{3000e2}} \omega _1^2 \omega _2+4 {g_{2020e2}} \omega _1^3 \omega
_2\right.\nonumber\\&&\left.+i {g_{1011e2}} {g_{1110}} \omega _1 \omega _2^2+i {g_{1011}}{g_{1110e2}} \omega _1 \omega _2^2+i {g_{0120e2}} {g_{2001}} \omega _1
\omega _2^2\right.\nonumber\\&&\left.+i {g_{0120}} {g_{2001e2}} \omega _1 \omega _2^2+i {g_{0021e2}} {g_{2100}} \omega _1 \omega _2^2+i {g_{0021}} {g_{2100e2}}
\omega _1 \omega _2^2\right.\nonumber\\&&\left.-3 i {g_{1020e2}} {g_{2010}} \omega _2^3-
3 i {g_{1020}} {g_{2010e2}} \omega _2^3-3 i {g_{0030e2}} {g_{3000}} \omega _2^3\right.\nonumber\\&&\left.-3 i{g_{0030}} {g_{3000e2}} \omega _2^3-{g_{2020e2}}
\omega _1 \omega _2^3\right],
\end{eqnarray}
\begin{eqnarray}
k_{1111} &=&\frac{1}{\omega _1 \left(\omega _1-2 \omega _2\right) \left(2 \omega _1-\omega _2\right) \omega _2 \left(2 \omega _1+\omega _2\right) \left(\omega
_1+2 \omega _2\right)}
\left[-8 i {g_{0201}} {g_{1011}} \omega _1^5\right.\nonumber\\&&\left.-8 i {g_{0102}} {g_{1110}} \omega _1^5-16 i {g_{0210}} {g_{1002}} \omega _1^4 \omega
_2+8 i {g_{1020}} {g_{1101}} \omega _1^4 \omega _2\right.\nonumber\\&&\left.+16 i {g_{0012}} {g_{1200}} \omega _1^4 \omega _2+8 i {g_{0120}} {g_{2001}} \omega
_1^4 \omega _2+8 i {g_{0111}} {g_{2010}} \omega _1^4 \omega _2\right.\nonumber\\&&\left.+8 i{g_{0021}} {g_{2100}} \omega _1^4 \omega _2+4 {g_{1111}} \omega _1^5
\omega _2+32 i {g_{0210}} {g_{1002}} \omega _1^3 \omega _2^2+\right.\nonumber\\&&\left.
34 i {g_{0201}} {g_{1011}} \omega _1^3 \omega _2^2+34 i {g_{0102}} {g_{1110}} \omega _1^3 \omega _2^2+32 i {g_{0012}} {g_{1200}}
\omega _1^3 \omega _2^2\right.\nonumber\\&&\left.-4 i {g_{0120}} {g_{2001}} \omega _1^3 \omega _2^2+4 i {g_{0021}} {g_{2100}} \omega _1^3 \omega _2^2+4 i {g_{0210}}
{g_{1002}} \omega _1^2 \omega _2^3\right.\nonumber\\&&\left.-34 i {g_{1020}} {g_{1101}} \omega _1^2 \omega _2^3-4 i{g_{0012}} {g_{1200}} \omega _1^2 \omega _2^3-32
i {g_{0120}} {g_{2001}} \omega _1^2 \omega _2^3\right.\nonumber\\&&\left.-
34 i {g_{0111}} {g_{2010}} \omega _1^2 \omega _2^3-32 i {g_{0021}} {g_{2100}} \omega _1^2 \omega _2^3-17 {g_{1111}} \omega _1^3
\omega _2^3-\right.\nonumber\\&&\left.8 i {g_{0210}} {g_{1002}} \omega _1 \omega _2^4-8 i {g_{0201}} {g_{1011}} \omega _1 \omega _2^4-8 i {g_{0102}} {g_{1110}}
\omega _1 \omega _2^4\right.\nonumber\\&&\left.-8 i {g_{0012}} {g_{1200}} \omega _1 \omega _2^4+16 i {g_{0120}} {g_{2001}} \omega _1 \omega _2^4-16 i {g_{0021}}
{g_{2100}} \omega _1 \omega _2^4\right.\nonumber\\&&\left.+8 i {g_{1020}} {g_{1101}} \omega _2^5+8 i {g_{0111}} {g_{2010}} \omega _2^5+4 {g_{1111}} \omega _1 \omega_2^5\right],
\end{eqnarray}
\begin{eqnarray}
k_{1111e1} &=&\frac{1}{\omega _1 \left(\omega _1-2 \omega _2\right) \left(2 \omega _1-\omega _2\right) \omega _2 \left(2 \omega _1+\omega _2\right) \left(\omega
_1+2 \omega _2\right)}\left[-8 i {g_{0201e1}} {g_{1011}} \omega _1^5\right.\nonumber\\&&\left.-8 i {g_{0201}} {g_{1011e1}} \omega _1^5-8 i {g_{0102e1}} {g_{1110}} \omega _1^5-8
i {g_{0102}} {g_{1110e1}} \omega _1^5\right.\nonumber\\&&\left.-16 i {g_{0210e1}} {g_{1002}} \omega _1^4 \omega _2-16 i {g_{0210}} {g_{1002e1}} \omega _1^4 \omega
_2+8 i {g_{1020e1}} {g_{1101}} \omega _1^4 \omega _2\right.\nonumber\\&&\left.+8 i {g_{1020}} {g_{1101e1}} \omega _1^4 \omega _2+16 i {g_{0012e1}} {g_{1200}} \omega
_1^4 \omega _2+
16 i {g_{0012}} {g_{1200e1}} \omega _1^4 \omega _2\right.\nonumber\\&&\left.+8 i {g_{0120e1}}{g_{2001}} \omega _1^4 \omega _2+8 i {g_{0120}}{g_{2001e1}}
\omega _1^4 \omega _2+8 i {g_{0111e1}} {g_{2010}} \omega _1^4 \omega _2\right.\nonumber\\&&\left.+8 i{g_{0111}} {g_{2010e1}} \omega _1^4 \omega _2+8 i {g_{0021e1}}
{g_{2100}} \omega _1^4 \omega _2+8 i {g_{0021}} {g_{2100e1}} \omega _1^4 \omega _2\right.\nonumber\\&&\left.+4 {g_{1111e1}} \omega _1^5 \omega _2+32 i{g_{0210e1}}
{g_{1002}} \omega _1^3 \omega _2^2+
32 i {g_{0210}} {g_{1002e1}} \omega _1^3 \omega _2^2+\right.\nonumber\\&&\left.34 i {g_{0201e1}} {g_{1011}} \omega _1^3 \omega _2^2+34 i {g_{0201}} {g_{1011e1}}
\omega _1^3 \omega _2^2+34 i {g_{0102e1}} {g_{1110}} \omega _1^3 \omega _2^2\right.\nonumber\\&&\left.+34 i {g_{0102}} {g_{1110e1}} \omega _1^3 \omega _2^2+32 i{g_{0012e1}}
{g_{1200}} \omega _1^3 \omega _2^2+32 i {g_{0012}} {g_{1200e1}} \omega _1^3 \omega _2^2\right.\nonumber\\&&\left.-4 i {g_{0120e1}} {g_{2001}} \omega _1^3 \omega
_2^2-4 i {g_{0120}} {g_{2001e1}} \omega _1^3 \omega _2^2+
4 i {g_{0021e1}} {g_{2100}} \omega _1^3 \omega _2^2\right.\nonumber\\&&\left.+4 i {g_{0021}} {g_{2100e1}} \omega _1^3 \omega _2^2+4 i {g_{0210e1}} {g_{1002}}
\omega _1^2 \omega _2^3+4 i {g_{0210}} {g_{1002e1}} \omega _1^2 \omega _2^3\right.\nonumber\\&&\left.-34 i {g_{1020e1}} {g_{1101}} \omega _1^2 \omega _2^3-34 i {g_{1020}}
{g_{1101e1}} \omega _1^2 \omega _2^3-4 i {g_{0012e1}} {g_{1200}} \omega _1^2 \omega _2^3\right.\nonumber\\&&\left.-4 i {g_{0012}}{g_{1200e1}} \omega _1^2 \omega
_2^3-32 i {g_{0120e1}} {g_{2001}} \omega _1^2 \omega _2^3-
32 i {g_{0120}} {g_{2001e1}} \omega _1^2 \omega _2^3\right.\nonumber\\&&\left.-34 i {g_{0111e1}} {g_{2010}} \omega _1^2 \omega _2^3-34 i {g_{0111}}{g_{2010e1}}
\omega _1^2 \omega _2^3-32 i {g_{0021e1}} {g_{2100}} \omega _1^2 \omega _2^3\right.\nonumber\\&&\left.-32 i {g_{0021}} {g_{2100e1}} \omega _1^2 \omega _2^3-17 {g_{1111e1}}
\omega _1^3 \omega _2^3-8 i {g_{0210e1}} {g_{1002}} \omega _1 \omega _2^4\right.\nonumber\\&&\left.-8 i {g_{0210}} {g_{1002e1}} \omega _1 \omega _2^4-8 i {g_{0201e1}}
{g_{1011}} \omega _1 \omega _2^4-
8 i {g_{0201}} {g_{1011e1}} \omega _1 \omega _2^4\right.\nonumber\\&&\left.-8 i {g_{0102e1}} {g_{1110}} \omega _1 \omega _2^4-8 i {g_{0102}} {g_{1110e1}}
\omega _1 \omega _2^4-8 i {g_{0012e1}} {g_{1200}} \omega _1 \omega _2^4\right.\nonumber\\&&\left.-8 i {g_{0012}} {g_{1200e1}} \omega _1 \omega _2^4+16 i {g_{0120e1}}
{g_{2001}} \omega _1 \omega _2^4+16 i {g_{0120}} {g_{2001e1}} \omega _1 \omega _2^4\right.\nonumber\\&&\left.-16 i {g_{0021e1}} {g_{2100}} \omega_1 \omega_2^4-16
i {g_{0021}} {g_{2100e1}} \omega _1 \omega _2^4+8 i {g_{1020e1}} {g_{1101}} \omega _2^5\right.\nonumber\\&&\left.+8 i {g_{1020}} {g_{1101e1}} \omega _2^5+8 i {g_{0111e1}} {g_{2010}} \omega _2^5+8i {g_{0111}} {g_{2010e1}} \omega_2^5\right.\nonumber\\&&\left.+4 {g_{1111e1}} \omega _1 \omega_2^5\right],
\end{eqnarray}
\begin{eqnarray}
k_{1111A} &=&\frac{1}{\omega _1 \left(\omega _1-2 \omega _2\right) \left(2 \omega _1-\omega _2\right) \omega _2 \left(2 \omega _1+\omega _2\right) \left(\omega
_1+2 \omega _2\right)}
\left[-8 i {g_{0201A}} {g_{1011}} \omega _1^5\right.\nonumber\\&&\left.-8 i  {g_{0201}}  {g_{1011A}} \omega _1^5-8 i  {g_{0102A}}  {g_{1110}} \omega _1^5-8 i
 {g_{0102}}  {g_{1110A}} \omega _1^5\right.\nonumber\\&&\left.-16 i  {g_{0210A}}  {g_{1002}} \omega _1^4 \omega _2-16 i  {g_{0210}}  {g_{1002A}} \omega _1^4 \omega _2+8
i  {g_{1020A}}  {g_{1101}} \omega _1^4 \omega _2\right.\nonumber\\&&\left.+8 i  {g_{1020}}  {g_{1101A}} \omega _1^4 \omega _2+16 i  {g_{0012A}}  {g_{1200}} \omega _1^4
\omega _2+16 i  {g_{0012}}  {g_{1200A}} \omega _1^4 \omega _2\right.\nonumber\\&&\left.+
8 i  {g_{0120A}}  {g_{2001}} \omega _1^4 \omega _2+8 i  {g_{0120}}  {g_{2001A}} \omega _1^4 \omega _2+8 i  {g_{0111A}}  {g_{2010}} \omega
_1^4 \omega _2\right.\nonumber\\&&\left.+8 i  {g_{0111}}  {g_{2010A}} \omega _1^4 \omega _2+8 i  {g_{0021A}}  {g_{2100}} \omega _1^4 \omega _2+8 i  {g_{0021}}  {g_{2100A}}
\omega _1^4 \omega _2\right.\nonumber\\&&\left.+4  {g_{1111A}} \omega _1^5 \omega _2+32 i  {g_{0210A}}  {g_{1002}} \omega _1^3 \omega _2^2+32 i  {g_{0210}}  {g_{1002A}}
\omega _1^3 \omega _2^2+\right.\nonumber\\&&\left.34 i  {g_{0201A}}  {g_{1011}} \omega _1^3 \omega _2^2+
34 i  {g_{0201}}  {g_{1011A}} \omega _1^3 \omega _2^2+34 i  {g_{0102A}}  {g_{1110}} \omega _1^3 \omega _2^2\right.\nonumber\\&&\left.+34 i  {g_{0102}}  {g_{1110A}}
\omega _1^3 \omega _2^2+32 i  {g_{0012A}}  {g_{1200}} \omega _1^3 \omega _2^2+32 i  {g_{0012}}  {g_{1200A}} \omega _1^3 \omega _2^2\right.\nonumber\\&&\left.-4 i  {g_{0120A}}
 {g_{2001}} \omega _1^3 \omega _2^2-4 i  {g_{0120}}  {g_{2001A}} \omega _1^3 \omega _2^2+4 i  {g_{0021A}}  {g_{2100}} \omega _1^3 \omega _2^2\right.\nonumber\\&&\left.+4
i  {g_{0021}}  {g_{2100A}} \omega _1^3 \omega _2^2+
4 i  {g_{0210A}}  {g_{1002}} \omega _1^2 \omega _2^3+4 i  {g_{0210}}  {g_{1002A}} \omega _1^2 \omega _2^3\right.\nonumber\\&&\left.-34 i  {g_{1020A}}  {g_{1101}}
\omega _1^2 \omega _2^3-34 i  {g_{1020}}  {g_{1101A}} \omega _1^2 \omega _2^3-4 i  {g_{0012A}}  {g_{1200}} \omega _1^2 \omega _2^3\right.\nonumber\\&&\left.-4 i  {g_{0012}}
 {g_{1200A}} \omega _1^2 \omega _2^3-32 i  {g_{0120A}}  {g_{2001}} \omega _1^2 \omega _2^3-32 i  {g_{0120}}  {g_{2001A}} \omega _1^2 \omega
_2^3\right.\nonumber\\&&\left.-34 i  {g_{0111A}}  {g_{2010}} \omega _1^2 \omega _2^3-
34 i  {g_{0111}}  {g_{2010A}} \omega _1^2 \omega _2^3-32 i  {g_{0021A}}  {g_{2100}} \omega _1^2 \omega _2^3\right.\nonumber\\&&\left.-32 i  {g_{0021}}  {g_{2100A}}
\omega _1^2 \omega _2^3-17  {g_{1111A}} \omega _1^3 \omega _2^3-8 i  {g_{0210A}}  {g_{1002}} \omega _1 \omega _2^4\right.\nonumber\\&&\left.-8 i  {g_{0210}}  {g_{1002A}}
\omega _1 \omega _2^4-8 i  {g_{0201A}}  {g_{1011}} \omega _1 \omega _2^4-8 i  {g_{0201}}  {g_{1011A}} \omega _1 \omega _2^4\right.\nonumber\\&&\left.-8 i  {g_{0102A}}
 {g_{1110}} \omega _1 \omega _2^4-
8 i  {g_{0102}}  {g_{1110A}} \omega _1 \omega _2^4-8 i  {g_{0012A}}  {g_{1200}} \omega _1 \omega _2^4\right.\nonumber\\&&\left.-8 i  {g_{0012}}  {g_{1200A}} \omega
_1 \omega _2^4+16 i  {g_{0120A}}  {g_{2001}} \omega _1 \omega _2^4+16 i  {g_{0120}}  {g_{2001A}} \omega _1 \omega _2^4\right.\nonumber\\&&\left.-16 i  {g_{0021A}}  {g_{2100}}
\omega _1 \omega _2^4-16 i  {g_{0021}}  {g_{2100A}} \omega _1 \omega _2^4+8 i  {g_{1020A}}  {g_{1101}} \omega _2^5\right.\nonumber\\&&\left.+8 i  {g_{1020}}  {g_{1101A}}
\omega _2^5+
8 i  {g_{0111A}}  {g_{2010}} \omega _2^5+8 i  {g_{0111}}  {g_{2010A}} \omega _2^5+4  {g_{1111A}} \omega _1 \omega _2^5\right],
\end{eqnarray}
\begin{eqnarray}
k_{1111e2} &=&\frac{1}{\omega _1 \left(\omega _1-2 \omega _2\right) \left(2 \omega _1-\omega _2\right) \omega _2 \left(2 \omega _1+\omega _2\right) \left(\omega
_1+2 \omega _2\right)}
\left[-8 i  {g_{0201e2}}  {g_{1011}} \omega _1^5\right.\nonumber\\&&\left.-8 i  {g_{0201}}  {g_{1011e2}} \omega _1^5-8 i  {g_{0102e2}}  {g_{1110}} \omega _1^5-8
i  {g_{0102}}  {g_{1110e2}} \omega _1^5\right.\nonumber\\&&\left.-16 i  {g_{0210e2}}  {g_{1002}} \omega _1^4 \omega _2-16 i  {g_{0210}}  {g_{1002e2}} \omega _1^4 \omega
_2+8 i  {g_{1020e2}}  {g_{1101}} \omega _1^4 \omega _2\right.\nonumber\\&&\left.+8 i  {g_{1020}}  {g_{1101e2}} \omega _1^4 \omega _2+16 i  {g_{0012e2}}  {g_{1200}} \omega
_1^4 \omega _2+
16 i  {g_{0012}}  {g_{1200e2}} \omega _1^4 \omega _2\right.\nonumber\\&&\left.+8 i  {g_{0120e2}}  {g_{2001}} \omega _1^4 \omega _2+8 i  {g_{0120}}  {g_{2001e2}}
\omega _1^4 \omega _2+8 i  {g_{0111e2}}  {g_{2010}} \omega _1^4 \omega _2\right.\nonumber\\&&\left.+8 i  {g_{0111}}  {g_{2010e2}} \omega _1^4 \omega _2+8 i  {g_{0021e2}}
 {g_{2100}} \omega _1^4 \omega _2+8 i  {g_{0021}}  {g_{2100e2}} \omega _1^4 \omega _2\right.\nonumber\\&&\left.+4  {g_{1111e2}} \omega _1^5 \omega _2+32 i  {g_{0210e2}}
 {g_{1002}} \omega _1^3 \omega _2^2+
32 i  {g_{0210}}  {g_{1002e2}} \omega _1^3 \omega _2^2+\right.\nonumber\\&&\left.34 i  {g_{0201e2}}  {g_{1011}} \omega _1^3 \omega _2^2+34 i  {g_{0201}}  {g_{1011e2}}
\omega _1^3 \omega _2^2+34 i  {g_{0102e2}}  {g_{1110}} \omega _1^3 \omega _2^2\right.\nonumber\\&&\left.+34 i  {g_{0102}}  {g_{1110e2}} \omega _1^3 \omega _2^2+32 i  {g_{0012e2}}
 {g_{1200}} \omega _1^3 \omega _2^2+32 i  {g_{0012}}  {g_{1200e2}} \omega _1^3 \omega _2^2\right.\nonumber\\&&\left.-4 i  {g_{0120e2}}  {g_{2001}} \omega _1^3 \omega
_2^2-4 i  {g_{0120}}  {g_{2001e2}} \omega _1^3 \omega _2^2+
4 i  {g_{0021e2}}  {g_{2100}} \omega _1^3 \omega _2^2\right.\nonumber\\&&\left.+4 i  {g_{0021}}  {g_{2100e2}} \omega _1^3 \omega _2^2+4 i  {g_{0210e2}}  {g_{1002}}
\omega _1^2 \omega _2^3+4 i  {g_{0210}}  {g_{1002e2}} \omega _1^2 \omega _2^3\right.\nonumber\\&&\left.-34 i  {g_{1020e2}}  {g_{1101}} \omega _1^2 \omega _2^3-34 i  {g_{1020}}
 {g_{1101e2}} \omega _1^2 \omega _2^3-4 i  {g_{0012e2}}  {g_{1200}} \omega _1^2 \omega _2^3\right.\nonumber\\&&\left.-4 i  {g_{0012}}  {g_{1200e2}} \omega _1^2 \omega
_2^3-32 i  {g_{0120e2}}  {g_{2001}} \omega _1^2 \omega _2^3-
32 i  {g_{0120}}  {g_{2001e2}} \omega _1^2 \omega _2^3\right.\nonumber\\&&\left.-34 i  {g_{0111e2}}  {g_{2010}} \omega _1^2 \omega _2^3-34 i  {g_{0111}}  {g_{2010e2}}
\omega _1^2 \omega _2^3-32 i  {g_{0021e2}}  {g_{2100}} \omega _1^2 \omega _2^3\right.\nonumber\\&&\left.-32 i  {g_{0021}}  {g_{2100e2}} \omega _1^2 \omega _2^3-17  {g_{1111e2}}
\omega _1^3 \omega _2^3-8 i  {g_{0210e2}}  {g_{1002}} \omega _1 \omega _2^4\right.\nonumber\\&&\left.-8 i  {g_{0210}}  {g_{1002e2}} \omega _1 \omega _2^4-8 i  {g_{0201e2}}
 {g_{1011}} \omega _1 \omega _2^4-
8 i  {g_{0201}}  {g_{1011e2}} \omega _1 \omega _2^4\right.\nonumber\\&&\left.-8 i  {g_{0102e2}}  {g_{1110}} \omega _1 \omega _2^4-8 i  {g_{0102}}  {g_{1110e2}}
\omega _1 \omega _2^4-8 i  {g_{0012e2}}  {g_{1200}} \omega _1 \omega _2^4\right.\nonumber\\&&\left.-8 i  {g_{0012}}  {g_{1200e2}} \omega _1 \omega _2^4+16 i  {g_{0120e2}}
 {g_{2001}} \omega _1 \omega _2^4+16 i  {g_{0120}}  {g_{2001e2}} \omega _1 \omega _2^4\right.\nonumber\\&&\left.-16 i  {g_{0021e2}}  {g_{2100}} \omega _1 \omega _2^4-16
i  {g_{0021}}  {g_{2100e2}} \omega _1 \omega _2^4+
8 i  {g_{1020e2}}  {g_{1101}} \omega _2^5\right.\nonumber\\&&\left.+8 i  {g_{1020}}  {g_{1101e2}} \omega _2^5+8 i  {g_{0111e2}}  {g_{2010}} \omega _2^5+8
i  {g_{0111}}  {g_{2010e2}} \omega _2^5\right.\nonumber\\&&\left.+4  {g_{1111e2}} \omega _1 \omega _2^5\right],
\end{eqnarray}
\begin{eqnarray}
k_{0202}&=&\frac{1}{\omega _1 \left(\omega _1-2 \omega _2\right) \omega _2 \left(\omega _1+2 \omega _2\right)}\left[-3 i  {g_{0102}}  {g_{0201}} \omega _1^3-3 i  {g_{0003}}  {g_{0300}} \omega _1^3\right.\nonumber\\&&\left.+i  {g_{0210}}  {g_{1002}}
\omega _1^2 \omega _2+i  {g_{0111}}  {g_{1101}} \omega _1^2 \omega _2+i  {g_{0012}}  {g_{1200}} \omega _1^2 \omega _2\right.\nonumber\\&&\left.+ {g_{0202}} \omega _1^3
\omega _2+12 i  {g_{0102}}  {g_{0201}} \omega _1 \omega _2^2+12 i  {g_{0003}}  {g_{0300}} \omega _1 \omega _2^2\right.\nonumber\\&&\left.-2 i  {g_{0210}}  {g_{1002}}
\omega _1 \omega _2^2+2 i  {g_{0012}}  {g_{1200}} \omega _1 \omega _2^2-4 i  {g_{0111}}  {g_{1101}} \omega _2^3-4  {g_{0202}} \omega _1 \omega
_2^3\right],
\end{eqnarray}
\begin{eqnarray}
k_{0202e1}&=&\frac{1}{\omega _1 \left(\omega _1-2 \omega _2\right) \omega _2 \left(\omega _1+2 \omega _2\right)}\left[-3 i  {g_{0102e1}}  {g_{0201}} \omega
_1^3-3 i  {g_{0102}}  {g_{0201e1}} \omega _1^3\right.\nonumber\\&&\left.-3 i  {g_{0003e1}}  {g_{0300}} \omega _1^3-3 i  {g_{0003}}  {g_{0300e1}} \omega _1^3+i  {g_{0210e1}}
 {g_{1002}} \omega _1^2 \omega _2\right.\nonumber\\&&\left.+i  {g_{0210}}  {g_{1002e1}} \omega _1^2 \omega _2+i  {g_{0111e1}}  {g_{1101}} \omega _1^2 \omega _2+i  {g_{0111}}
 {g_{1101e1}} \omega _1^2 \omega _2\right.\nonumber\\&&\left.+i  {g_{0012e1}}  {g_{1200}} \omega _1^2 \omega _2+
i  {g_{0012}}  {g_{1200e1}} \omega _1^2 \omega _2+ {g_{0202e1}} \omega _1^3 \omega _2\right.\nonumber\\&&\left.+12 i  {g_{0102e1}}  {g_{0201}} \omega _1 \omega _2^2+12
i  {g_{0102}}  {g_{0201e1}} \omega _1 \omega _2^2+12 i  {g_{0003e1}}  {g_{0300}} \omega _1 \omega _2^2\right.\nonumber\\&&\left.+12 i  {g_{0003}}  {g_{0300e1}} \omega
_1 \omega _2^2-2 i  {g_{0210e1}}  {g_{1002}} \omega _1 \omega _2^2-2 i  {g_{0210}}  {g_{1002e1}} \omega _1 \omega _2^2\right.\nonumber\\&&\left.+2 i  {g_{0012e1}}  {g_{1200}}
\omega _1 \omega _2^2+
2 i  {g_{0012}}  {g_{1200e1}} \omega _1 \omega _2^2-4 i  {g_{0111e1}}  {g_{1101}} \omega _2^3\right.\nonumber\\&&\left.-4 i  {g_{0111}}  {g_{1101e1}} \omega
_2^3-4  {g_{0202e1}} \omega _1 \omega _2^3\right],
\end{eqnarray}
\begin{eqnarray}
k_{0202A}&=&\frac{1}{\omega _1 \left(\omega _1-2 \omega _2\right) \omega _2 \left(\omega _1+2 \omega _2\right)}\left[-3 i  {g_{0102A}}  {g_{0201}} \omega
_1^3-3 i  {g_{0102}}  {g_{0201A}} \omega _1^3\right.\nonumber\\&&\left.-3 i  {g_{0003A}}  {g_{0300}} \omega _1^3-3 i  {g_{0003}}  {g_{0300A}} \omega _1^3+i  {g_{0210A}}
 {g_{1002}} \omega _1^2 \omega _2\right.\nonumber\\&&\left.+i  {g_{0210}}  {g_{1002A}} \omega _1^2 \omega _2+i  {g_{0111A}}  {g_{1101}} \omega _1^2 \omega _2+i  {g_{0111}}
 {g_{1101A}} \omega _1^2 \omega _2\right.\nonumber\\&&\left.+i  {g_{0012A}}  {g_{1200}} \omega _1^2 \omega _2+
i  {g_{0012}}  {g_{1200A}} \omega _1^2 \omega _2+ {g_{0202A}} \omega _1^3 \omega _2\right.\nonumber\\&&\left.+12 i  {g_{0102A}}  {g_{0201}} \omega _1 \omega _2^2+12
i  {g_{0102}}  {g_{0201A}} \omega _1 \omega _2^2+12 i  {g_{0003A}}  {g_{0300}} \omega _1 \omega _2^2\right.\nonumber\\&&\left.+12 i  {g_{0003}}  {g_{0300A}} \omega _1
\omega _2^2-2 i  {g_{0210A}}  {g_{1002}} \omega _1 \omega _2^2-2 i  {g_{0210}}  {g_{1002A}} \omega _1 \omega _2^2\right.\nonumber\\&&\left.+2 i  {g_{0012A}}  {g_{1200}}
\omega _1 \omega _2^2+2 i  {g_{0012}}  {g_{1200A}} \omega _1 \omega _2^2-
4 i  {g_{0111A}}  {g_{1101}} \omega _2^3\right.\nonumber\\&&\left.-4 i  {g_{0111}}  {g_{1101A}} \omega _2^3-4  {g_{0202A}} \omega _1 \omega _2^3\right],
\end{eqnarray}
\begin{eqnarray}
k_{0202e2}&=&\frac{1}{\omega _1 \left(\omega _1-2 \omega _2\right) \omega _2 \left(\omega _1+2 \omega _2\right)}\left[-3 i  {g_{0102e2}}  {g_{0201}} \omega
_1^3-3 i  {g_{0102}}  {g_{0201e2}} \omega _1^3\right.\nonumber\\&&\left.-3 i  {g_{0003e2}}  {g_{0300}} \omega _1^3-3 i  {g_{0003}}  {g_{0300e2}} \omega _1^3+i  {g_{0210e2}}
 {g_{1002}} \omega _1^2 \omega _2\right.\nonumber\\&&\left.+i  {g_{0210}}  {g_{1002e2}} \omega _1^2 \omega _2+i  {g_{0111e2}}  {g_{1101}} \omega _1^2 \omega _2+i  {g_{0111}}
 {g_{1101e2}} \omega _1^2 \omega _2\right.\nonumber\\&&\left.+i  {g_{0012e2}}  {g_{1200}} \omega _1^2 \omega _2+
i  {g_{0012}}  {g_{1200e2}} \omega _1^2 \omega _2+ {g_{0202e2}} \omega _1^3 \omega _2\right.\nonumber\\&&\left.+12 i  {g_{0102e2}}  {g_{0201}} \omega _1 \omega _2^2+12
i  {g_{0102}}  {g_{0201e2}} \omega _1 \omega _2^2+12 i  {g_{0003e2}}  {g_{0300}} \omega _1 \omega _2^2\right.\nonumber\\&&\left.+12 i  {g_{0003}}  {g_{0300e2}} \omega
_1 \omega _2^2-2 i  {g_{0210e2}}  {g_{1002}} \omega _1 \omega _2^2-2 i  {g_{0210}}  {g_{1002e2}} \omega _1 \omega _2^2\right.\nonumber\\&&\left.+2 i  {g_{0012e2}}  {g_{1200}}
\omega _1 \omega _2^2+
2 i  {g_{0012}}  {g_{1200e2}} \omega _1 \omega _2^2-4 i  {g_{0111e2}}  {g_{1101}} \omega _2^3\right.\nonumber\\&&\left.-4 i  {g_{0111}}  {g_{1101e2}} \omega
_2^3-4  {g_{0202e2}} \omega _1 \omega _2^3\right].\end{eqnarray}
\subsection{Coefficient in $D_2$}
\label{subsec:coefD2}
\begin{eqnarray}D_{20}&=&\frac{-i}{4 \omega _1^5 \omega _2-17 \omega _1^3 \omega 
_2^3+4 \omega _1 \omega _2^5}
\left[-12 \left( {g_{1020}}  {g_{2010}}+ {g_{0030}}  {g_{3000}}\right) \omega _2^7+4 \omega _1^7 \left(3  {g_{0102}}  {g_{0201}}\right.\right.\nonumber\\&&\left.\left.+3  {g_{0003}}  {g_{0300}}+i \omega _2{g_{0202}}\right)-
4 \omega _1^6 \omega _2 \left( {g_{0210}}  {g_{1002}}-2  {g_{0201}} 
 {g_{1011}}+ {g_{0111}}  {g_{1101}}\right.\right.\nonumber\\&&\left.\left.-2  {g_{0102}}  {g_{1110}}
 +{g_{0012}}  {g_{1200}}-i  \omega _2 {g_{1111}}\right)+
\omega _1^2 \omega _2^5 \left(8  {g_{0210}}  {g_{1002}}+8  {g_{0201}} 
 {g_{1011}}\right.\right.\nonumber\\&&\left.\left.-4  {g_{0111}}  {g_{1101}}+8  {g_{0102}}  {g_{1110}}
+8  {g_{0012}}  {g_{1200}}-24  {g_{0120}}  {g_{2001}}+51  {g_{1020}} 
 {g_{2010}}\right.\right.\nonumber\\&&\left.\left.+24  {g_{0021}}  {g_{2100}}+51  {g_{0030}}  {g_{3000}}+4 i \omega _2 {g_{1111}} \right)-\omega _1^4 \omega _2^3 \left(31  {g_{0210}}  {g_{1002}}+34  {g_{0201}} 
 {g_{1011}}\right.\right.\nonumber\\&&\left.\left.-17  {g_{0111}}  {g_{1101}}+34  {g_{0102}}  {g_{1110}}+31  {g_{0012}}  {g_{1200}}-6  {g_{0120}}  {g_{2001}}+12  {g_{1020}} 
 {g_{2010}}\right.\right.\nonumber\\&&\left.\left.+6  {g_{0021}}  {g_{2100}}+12  {g_{0030}}  {g_{3000}}+17 i \omega _2 {g_{1111}} \right)+\omega _1^3 \omega _2^4 \left(12  {g_{0102}}  {g_{0201}}+12  {g_{0003}} 
 {g_{0300}}\right.\right.\nonumber\\&&\left.\left.-6  {g_{0210}}  {g_{1002}}+34  {g_{1020}}  {g_{1101}}
-17  {g_{1011}}  {g_{1110}}+6  {g_{0012}}  {g_{1200}}+31  {g_{0120}} 
 {g_{2001}}\right.\right.\nonumber\\&&\left.\left.+34  {g_{0111}}  {g_{2010}}+31  {g_{0021}}  {g_{2100}}+i\omega _2 \left(4  {g_{0202}}-17  {g_{2020}}\right)\right)-\omega _1^5 \omega _2^2 \left(51  {g_{0102}}  {g_{0201}}\right.\right.\nonumber\\&&\left.\left.+51  {g_{0003}} {g_{0300}}-24  {g_{0210}}  {g_{1002}}+8  {g_{1020}}  {g_{1101}}
-4  {g_{1011}}  {g_{1110}}+24  {g_{0012}}  {g_{1200}}\right.\right.\nonumber\\&&\left.\left.+8  {g_{0120}} 
 {g_{2001}}+8  {g_{0111}}  {g_{2010}}+8  {g_{0021}}  {g_{2100}}+i \left(17  {g_{0202}}-4  {g_{2020}}\right) \omega _2\right)+4 \omega _1 \omega _2^6\right.\nonumber\\&&\left. \left(-2  {g_{1020}}  {g_{1101}}+ {g_{1011}} 
 {g_{1110}}+ {g_{0120}}  {g_{2001}}-2  {g_{0111}}  {g_{2010}}
+ {g_{0021}}  {g_{2100}}+i  \omega_2 {g_{2020}}\right)\right],
\end{eqnarray}
\begin{eqnarray}
D_{21}&=&\frac{-i}{4 \omega _1^5 \omega _2-17 \omega _1^3 
\omega _2^3+4 \omega _1 \omega _2^5} 
\left[-12 \left( {g_{1020e1}}  {g_{2010}}+ {g_{1020}} 
 {g_{2010e1}}+ {g_{0030e1}}  {g_{3000}}\right.\right.\nonumber\\&&\left.\left.+ {g_{0030}}  {g_{3000e1}}\right) \omega 
_2^7
+4 \omega _1^7 \left(3 \left( {g_{0102e1}}  {g_{0201}}+ {g_{0102}} 
 {g_{0201e1}}+ {g_{0003e1}}  {g_{0300}}+ {g_{0003}}  {g_{0300e1}}\right)\right.\right.\nonumber\\&&\left.\left.
+i  \omega _2 {g_{0202e1}}\right)-
4 \omega _1^6 \omega _2 \left( {g_{0210e1}}  {g_{1002}}+ {g_{0210}} 
 {g_{1002e1}}-2  {g_{0201e1}}  {g_{1011}}-2  {g_{0201}}  {g_{1011e1}}\right.\right.\nonumber\\&&\left.\left.+
 {g_{0111e1}}  {g_{1101}}+ {g_{0111}}  {g_{1101e1}}-2  {g_{0102e1}} 
 {g_{1110}}-2  {g_{0102}}  {g_{1110e1}}+
 {g_{0012e1}}  {g_{1200}}\right.\right.\nonumber\\&&\left.\left.+ {g_{0012}}  {g_{1200e1}}-i  \omega _2{g_{1111e1}} 
\right)
+\omega _1^2 \omega _2^5 \left(8  {g_{0210e1}}  {g_{1002}}+8  {g_{0210}} 
 {g_{1002e1}}+8  {g_{0201e1}}  {g_{1011}}\right.\right.\nonumber\\&&\left.\left.+8  {g_{0201}}  {g_{1011e1}}-
4  {g_{0111e1}}  {g_{1101}}-4  {g_{0111}}  {g_{1101e1}}+8  {g_{0102e1}} 
 {g_{1110}}+8  {g_{0102}}  {g_{1110e1}}\right.\right.\nonumber\\&&\left.\left.+
8  {g_{0012e1}}  {g_{1200}}+8  {g_{0012}}  {g_{1200e1}}-24  {g_{0120e1}} 
 {g_{2001}}-24  {g_{0120}}  {g_{2001e1}}+
51  {g_{1020e1}}  {g_{2010}}\right.\right.\nonumber\\&&\left.\left.+51  {g_{1020}}  {g_{2010e1}}+24  {g_{0021e1}} 
 {g_{2100}}+24  {g_{0021}}  {g_{2100e1}}
+51  {g_{0030e1}}  {g_{3000}}+51  {g_{0030}}  {g_{3000e1}}\right.\right.\nonumber\\&&\left.\left.+4 i 
 {g_{1111e1}} \omega _2\right)-
\omega _1^4 \omega _2^3 \left(31  {g_{0210e1}}  {g_{1002}}+31  {g_{0210}} 
 {g_{1002e1}}+34  {g_{0201e1}}  {g_{1011}}\right.\right.\nonumber\\&&\left.\left.+34  {g_{0201}}  {g_{1011e1}}-
17  {g_{0111e1}}  {g_{1101}}-17  {g_{0111}}  {g_{1101e1}}+34  {g_{0102e1}} 
 {g_{1110}}\right.\right.\nonumber\\&&\left.\left.+34  {g_{0102}}  {g_{1110e1}}+
31  {g_{0012e1}}  {g_{1200}}+31  {g_{0012}}  {g_{1200e1}}-6  {g_{0120e1}} 
 {g_{2001}}-6  {g_{0120}}  {g_{2001e1}}\right.\right.\nonumber\\&&\left.\left.+
12  {g_{1020e1}}  {g_{2010}}+12  {g_{1020}}  {g_{2010e1}}+6  {g_{0021e1}} 
 {g_{2100}}+6  {g_{0021}}  {g_{2100e1}}+
12  {g_{0030e1}}  {g_{3000}}\right.\right.\nonumber\\&&\left.\left.+12  {g_{0030}}  {g_{3000e1}}+17 i  \omega _2{g_{1111e1}}\right)+
\omega _1^3 \omega _2^4 \left(12  {g_{0102e1}}  {g_{0201}}+12  {g_{0102}} 
 {g_{0201e1}}\right.\right.\nonumber\\&&\left.\left.+12  {g_{0003e1}}  {g_{0300}}+12  {g_{0003}}  {g_{0300e1}}-
6  {g_{0210e1}}  {g_{1002}}-6  {g_{0210}}  {g_{1002e1}}+34  {g_{1020e1}} 
 {g_{1101}}\right.\right.\nonumber\\&&\left.\left.+34  {g_{1020}}  {g_{1101e1}}-
17  {g_{1011e1}}  {g_{1110}}-17  {g_{1011}}  {g_{1110e1}}+6  {g_{0012e1}} 
 {g_{1200}}+6  {g_{0012}}  {g_{1200e1}}\right.\right.\nonumber\\&&\left.\left.+
31  {g_{0120e1}}  {g_{2001}}+31  {g_{0120}}  {g_{2001e1}}+34  {g_{0111e1}} 
 {g_{2010}}+34  {g_{0111}}  {g_{2010e1}}+
31  {g_{0021e1}}  {g_{2100}}\right.\right.\nonumber\\&&\left.\left.+31  {g_{0021}}  {g_{2100e1}}+i \omega _2\left(4 
 {g_{0202e1}}-17  {g_{2020e1}}\right) \right)-
\omega _1^5 \omega _2^2 \left(51  {g_{0102e1}}  {g_{0201}}\right.\right.\nonumber\\&&\left.\left.+51  {g_{0102}} 
 {g_{0201e1}}+51  {g_{0003e1}}  {g_{0300}}+51  {g_{0003}}  {g_{0300e1}}-
24  {g_{0210e1}}  {g_{1002}}\right.\right.\nonumber\\&&\left.\left.-24  {g_{0210}}  {g_{1002e1}}+8  {g_{1020e1}} 
 {g_{1101}}+8  {g_{1020}}  {g_{1101e1}}-
4  {g_{1011e1}}  {g_{1110}}-4  {g_{1011}}  {g_{1110e1}}\right.\right.\nonumber\\&&\left.\left.+24  {g_{0012e1}} 
 {g_{1200}}+24  {g_{0012}}  {g_{1200e1}}+
8  {g_{0120e1}}  {g_{2001}}+8  {g_{0120}}  {g_{2001e1}}+8  {g_{0111e1}} 
 {g_{2010}}\right.\right.\nonumber\\&&\left.\left.+8  {g_{0111}}  {g_{2010e1}}+
8  {g_{0021e1}}  {g_{2100}}+8  {g_{0021}}  {g_{2100e1}}+i \omega _2\left(17 
 {g_{0202e1}}-4  {g_{2020e1}}\right) \right)\right.\nonumber\\&&\left.+
4 \omega _1 \omega _2^6 \left(-2  {g_{1020e1}}  {g_{1101}}-2  {g_{1020}} 
 {g_{1101e1}}+ {g_{1011e1}}  {g_{1110}}+ {g_{1011}}  {g_{1110e1}}+
 {g_{0120e1}}  {g_{2001}}\right.\right.\nonumber\\&&\left.\left.+ {g_{0120}}  {g_{2001e1}}-2  {g_{0111e1}} 
 {g_{2010}}-2  {g_{0111}}  {g_{2010e1}}+
{g_{0021e1}}  {g_{2100}}+ {g_{0021}}  {g_{2100e1}}\right.\right.\nonumber\\&&\left.\left.+i 
 {g_{2020e1}} \omega _2\right)\right],
\end{eqnarray}

\begin{eqnarray}
D_{22}&=&\frac{-i}{4 \omega _1^5 \omega _2-17 \omega _1^3 \omega 
_2^3+4 \omega _1 \omega _2^5}
\left[-12 \left( {g_{1020A}}  {g_{2010}}+ {g_{1020}}  {g_{2010A}}+ {g_{0030A}} 
 {g_{3000}}\right.\right.\nonumber\\&&\left.\left.+ {g_{0030}}  {g_{3000A}}\right) \omega _2^7+
4 \omega _1^7 \left(3 \left( {g_{0102A}}  {g_{0201}}+ {g_{0102}} 
 {g_{0201A}}+ {g_{0003A}}  {g_{0300}}+ {g_{0003}}  {g_{0300A}}\right)\right.\right.\nonumber\\&&\left.\left.+i 
 \omega _2{g_{0202A}}\right)-4 \omega _1^6 \omega _2 \left( {g_{0210A}}  {g_{1002}}+ {g_{0210}}  {g_{1002A}}-2 
 {g_{0201A}}  {g_{1011}}-2  {g_{0201}}  {g_{1011A}}
 \right.\right.\nonumber\\&&\left.\left.+{g_{0111A}}  {g_{1101}}+ {g_{0111}}  {g_{1101A}}-2  {g_{0102A}} {g_{1110}}-2  {g_{0102}}  {g_{1110A}}+
{g_{0012A}}  {g_{1200}}\right.\right.\nonumber\\&&\left.\left.+ {g_{0012}}  {g_{1200A}}-i \omega _2 {g_{1111A}} 
\right)+\omega _1^2 \omega _2^5 \left(8  {g_{0210A}}  {g_{1002}}+8  {g_{0210}} 
 {g_{1002A}}+8  {g_{0201A}}  {g_{1011}}\right.\right.\nonumber\\&&\left.\left.+8  {g_{0201}}  {g_{1011A}}-4  {g_{0111A}}  {g_{1101}}-4  {g_{0111}}  {g_{1101A}}+8  {g_{0102A}} {g_{1110}}+8  {g_{0102}}  {g_{1110A}}
\right.\right.\nonumber\\&&\left.\left.+8  {g_{0012A}}  {g_{1200}}+8  {g_{0012}}  {g_{1200A}}-24  {g_{0120A}} 
 {g_{2001}}-24  {g_{0120}}  {g_{2001A}}+51  {g_{1020A}}  {g_{2010}}\right.\right.\nonumber\\&&\left.\left.+51  {g_{1020}}  {g_{2010A}}+24  {g_{0021A}} {g_{2100}}+24  {g_{0021}}  {g_{2100A}}
+51  {g_{0030A}}  {g_{3000}}+51  {g_{0030}}  {g_{3000A}}\right.\right.\nonumber\\&&\left.\left.+4 i \omega _2 
 {g_{1111A}}\right)-\omega _1^4 \omega _2^3 \left(31  {g_{0210A}}  {g_{1002}}+31  {g_{0210}} 
 {g_{1002A}}+34  {g_{0201A}}  {g_{1011}}\right.\right.\nonumber\\&&\left.\left.+34  {g_{0201}}  {g_{1011A}}
-17  {g_{0111A}}  {g_{1101}}-17  {g_{0111}}  {g_{1101A}}+34  {g_{0102A}} 
 {g_{1110}}+34  {g_{0102}}  {g_{1110A}}\right.\right.\nonumber\\&&\left.\left.+31  {g_{0012A}}  {g_{1200}}+31  {g_{0012}}  {g_{1200A}}-6  {g_{0120A}} 
 {g_{2001}}-6  {g_{0120}}  {g_{2001A}}+
12  {g_{1020A}}  {g_{2010}}\right.\right.\nonumber\\&&\left.\left.+12  {g_{1020}}  {g_{2010A}}+6  {g_{0021A}} {g_{2100}}+6  {g_{0021}}  {g_{2100A}}+
12  {g_{0030A}}  {g_{3000}}+12  {g_{0030}}  {g_{3000A}}\right.\right.\nonumber\\&&\left.\left.+17 i \omega _2 {g_{1111A}}\right)
+\omega _1^3 \omega _2^4\left (12  {g_{0102A}}  {g_{0201}}+12  {g_{0102}} 
 {g_{0201A}}+12  {g_{0003A}}  {g_{0300}}\right.\right.\nonumber\\&&\left.\left.+12  {g_{0003}}  {g_{0300A}}-6  {g_{0210A}}  {g_{1002}}-6  {g_{0210}}  {g_{1002A}}+34  {g_{1020A}} {g_{1101}}+34  {g_{1020}}  {g_{1101A}}
\right.\right.\nonumber\\&&\left.\left.-17  {g_{1011A}}  {g_{1110}}-17  {g_{1011}}  {g_{1110A}}+6  {g_{0012A}} {g_{1200}}+6  {g_{0012}}  {g_{1200A}}+
31  {g_{0120A}}  {g_{2001}}\right.\right.\nonumber\\&&\left.\left.+31  {g_{0120}}  {g_{2001A}}+34  {g_{0111A}} 
 {g_{2010}}+34  {g_{0111}}  {g_{2010A}}
+31  {g_{0021A}}  {g_{2100}}+31  {g_{0021}}  {g_{2100A}}\right.\right.\nonumber\\&&\left.\left.+i \omega _2 \left(4 {g_{0202A}}-17 {g_{2020A}}\right)\right)-
\omega _1^5 \omega _2^2 \left(51  {g_{0102A}}  {g_{0201}}+51  {g_{0102}} 
 {g_{0201A}}+51  {g_{0003A}}  {g_{0300}}\right.\right.\nonumber\\&&\left.\left.+51  {g_{0003}}  {g_{0300A}}-
24  {g_{0210A}}  {g_{1002}}-24  {g_{0210}}  {g_{1002A}}+8  {g_{1020A}} {g_{1101}}+8  {g_{1020}}  {g_{1101A}}\right.\right.\nonumber\\&&\left.\left.-
4  {g_{1011A}}  {g_{1110}}-4  {g_{1011}}  {g_{1110A}}+24  {g_{0012A}} 
 {g_{1200}}+24  {g_{0012}}  {g_{1200A}}
+8  {g_{0120A}}  {g_{2001}}\right.\right.\nonumber\\&&\left.\left.+8  {g_{0120}}  {g_{2001A}}+8  {g_{0111A}} {g_{2010}}+8  {g_{0111}}  {g_{2010A}}+8  {g_{0021A}}  {g_{2100}}+8  {g_{0021}}  {g_{2100A}}\right.\right.\nonumber\\&&\left.\left.+i  \omega _2\left(17 
 {g_{0202A}}-4  {g_{2020A}}\right)\right)+4 \omega _1 \omega _2^6 \left(-2  {g_{1020A}}  {g_{1101}}-2  {g_{1020}} {g_{1101A}}+ {g_{1011A}}  {g_{1110}}\right.\right.\nonumber\\&&\left.\left.+ {g_{1011}}  {g_{1110A}}+{g_{0120A}}  {g_{2001}}+ {g_{0120}}  {g_{2001A}}-2  {g_{0111A}} 
 {g_{2010}}-2  {g_{0111}}  {g_{2010A}}\right.\right.\nonumber\\&&\left.\left.
+{g_{0021A}}  {g_{2100}}+ {g_{0021}}  {g_{2100A}}+i \omega _2 {g_{2020A}} \right)\right],
\end{eqnarray}

\begin{eqnarray}
D_{23}&=&\frac{-i}{4 \omega _1^5 \omega _2-17 \omega _1^3 
\omega _2^3+4 \omega _1 \omega _2^5}
\left[-12 \left( {g_{1020e2}}  {g_{2010}}+ {g_{1020}} 
 {g_{2010e2}}+ {g_{0030e2}}  {g_{3000}}\right.\right.\nonumber\\&&\left.\left.+ {g_{0030}}  {g_{3000e2}}\right) \omega 
_2^7+4 \omega _1^7 \left(3 \left( {g_{0102e2}}  {g_{0201}}+ {g_{0102}} 
 {g_{0201e2}}+ {g_{0003e2}}  {g_{0300}}\right.\right.\right.\nonumber\\&&\left.\left.\left.+ {g_{0003}}  {g_{0300e2}}\right)+
i  \omega _2{g_{0202e2}} \right)-4 \omega _1^6 \omega _2 \left( {g_{0210e2}}  {g_{1002}}+ {g_{0210}} 
 {g_{1002e2}}-2  {g_{0201e2}}  {g_{1011}}\right.\right.\nonumber\\&&\left.\left.-2  {g_{0201}}  {g_{1011e2}}+
 {g_{0111e2}}  {g_{1101}}+ {g_{0111}}  {g_{1101e2}}-2  {g_{0102e2}} 
 {g_{1110}}-2  {g_{0102}}  {g_{1110e2}}+\right.\right.\nonumber\\&&\left.\left.
 {g_{0012e2}}  {g_{1200}}+ {g_{0012}}  {g_{1200e2}}-i \omega _2 {g_{1111e2}} 
\right)+\omega _1^2 \omega _2^5 \left(8  {g_{0210e2}}  {g_{1002}}+8  {g_{0210}} 
 {g_{1002e2}}\right.\right.\nonumber\\&&\left.\left.+8  {g_{0201e2}}  {g_{1011}}+8  {g_{0201}}  {g_{1011e2}}-
4  {g_{0111e2}}  {g_{1101}}-4  {g_{0111}}  {g_{1101e2}}+8  {g_{0102e2}} 
 {g_{1110}}\right.\right.\nonumber\\&&\left.\left.+8  {g_{0102}}  {g_{1110e2}}+8  {g_{0012e2}}  {g_{1200}}+8  {g_{0012}}  {g_{1200e2}}-24  {g_{0120e2}} 
 {g_{2001}}-24  {g_{0120}}  {g_{2001e2}}\right.\right.\nonumber\\&&\left.\left.+
51  {g_{1020e2}}  {g_{2010}}+51  {g_{1020}}  {g_{2010e2}}+24  {g_{0021e2}} 
 {g_{2100}}+24  {g_{0021}}  {g_{2100e2}}
\right.\right.\nonumber\\&&\left.\left.+51  {g_{0030e2}}  {g_{3000}}+51  {g_{0030}}  {g_{3000e2}}+4 i  \omega _2{g_{1111e2}}\right)-
\omega _1^4 \omega _2^3 \left(31  {g_{0210e2}}  {g_{1002}}\right.\right.\nonumber\\&&\left.\left.+31  {g_{0210}} 
 {g_{1002e2}}+34  {g_{0201e2}}  {g_{1011}}+34  {g_{0201}}  {g_{1011e2}}-
17  {g_{0111e2}}  {g_{1101}}\right.\right.\nonumber\\&&\left.\left.-17  {g_{0111}}  {g_{1101e2}}+34  {g_{0102e2}} 
 {g_{1110}}+34  {g_{0102}}  {g_{1110e2}}+
31  {g_{0012e2}}  {g_{1200}}+31  {g_{0012}}  {g_{1200e2}}\right.\right.\nonumber\\&&\left.\left.-6  {g_{0120e2}} 
 {g_{2001}}-6  {g_{0120}}  {g_{2001e2}}+
12  {g_{1020e2}}  {g_{2010}}+12  {g_{1020}}  {g_{2010e2}}+6  {g_{0021e2}} 
 {g_{2100}}\right.\right.\nonumber\\&&\left.\left.+6  {g_{0021}}  {g_{2100e2}}+
12  {g_{0030e2}}  {g_{3000}}+12  {g_{0030}}  {g_{3000e2}}+17 i\omega _2 
 {g_{1111e2}} \right)\right.\nonumber\\&&\left.+\omega _1^3 \omega _2^4 \left(12  {g_{0102e2}}  {g_{0201}}+12  {g_{0102}} 
 {g_{0201e2}}+12  {g_{0003e2}}  {g_{0300}}+12  {g_{0003}}  {g_{0300e2}}\right.\right.\nonumber\\&&\left.\left.-
6  {g_{0210e2}}  {g_{1002}}-6  {g_{0210}}  {g_{1002e2}}+34  {g_{1020e2}} 
 {g_{1101}}+34  {g_{1020}}  {g_{1101e2}}-
17  {g_{1011e2}}  {g_{1110}}\right.\right.\nonumber\\&&\left.\left.-17  {g_{1011}}  {g_{1110e2}}+6  {g_{0012e2}} 
 {g_{1200}}+6  {g_{0012}}  {g_{1200e2}}+
31  {g_{0120e2}}  {g_{2001}}+31  {g_{0120}}  {g_{2001e2}}\right.\right.\nonumber\\&&\left.\left.+34  {g_{0111e2}} 
 {g_{2010}}+34  {g_{0111}}  {g_{2010e2}}+
31  {g_{0021e2}}  {g_{2100}}+31  {g_{0021}}  {g_{2100e2}}\right.\right.\nonumber\\&&\left.\left.+i  \omega _2\left(4 
 {g_{0202e2}}-17  {g_{2020e2}}\right)\right)-
\omega _1^5 \omega _2^2 \left(51  {g_{0102e2}}  {g_{0201}}+51  {g_{0102}} 
 {g_{0201e2}}\right.\right.\nonumber\\&&\left.\left.+51  {g_{0003e2}}  {g_{0300}}+51  {g_{0003}}  {g_{0300e2}}
24  {g_{0210e2}}  {g_{1002}}-24  {g_{0210}}  {g_{1002e2}}+8  {g_{1020e2}} 
 {g_{1101}}\right.\right.\nonumber\\&&\left.\left.+8  {g_{1020}}  {g_{1101e2}}-
4  {g_{1011e2}}  {g_{1110}}-4  {g_{1011}}  {g_{1110e2}}+24  {g_{0012e2}} 
 {g_{1200}}+24  {g_{0012}}  {g_{1200e2}}\right.\right.\nonumber\\&&\left.\left.+
8  {g_{0120e2}}  {g_{2001}}+8  {g_{0120}}  {g_{2001e2}}+8  {g_{0111e2}} 
 {g_{2010}}+8  {g_{0111}}  {g_{2010e2}}+
8  {g_{0021e2}}  {g_{2100}}\right.\right.\nonumber\\&&\left.\left.+8  {g_{0021}}  {g_{2100e2}}+i  \omega _2\left(17 
 {g_{0202e2}}-4  {g_{2020e2}}\right)\right)+
4 \omega _1 \omega _2^6 \left(-2  {g_{1020e2}}  {g_{1101}}-2  {g_{1020}} 
 {g_{1101e2}}\right.\right.\nonumber\\&&\left.\left.+ {g_{1011e2}}  {g_{1110}}+ {g_{1011}}  {g_{1110e2}}+
 {g_{0120e2}}  {g_{2001}}+ {g_{0120}}  {g_{2001e2}}\right.\right.\nonumber\\&&\left.\left.-2  {g_{0111e2}} 
 {g_{2010}}-2  {g_{0111}}  {g_{2010e2}}+
 {g_{0021e2}}  {g_{2100}}+ {g_{0021}}  {g_{2100e2}}+i \omega _2 
 {g_{2020e2}}\right)\right].\end{eqnarray}
\bibliographystyle{spbasic} 
 \bibliography{ref}
\end{document}